\documentclass[fleqn,usenatbib]{mnras}

\usepackage[T1]{fontenc}

\DeclareRobustCommand{\VAN}[3]{#2}
\let\VANthebibliography\thebibliography
\def\thebibliography{\DeclareRobustCommand{\VAN}[3]{##3}\VANthebibliography}

\usepackage{graphicx}	
\usepackage{amsmath}	
\usepackage{amssymb}	
\usepackage{natbib}
\usepackage{nccmath}
\usepackage{siunitx}
\usepackage{xcolor} 

\newcommand{\hst}{{\it HST}}
\newcommand{\HST}{{\it HST}}

\newcommand{\lenstool}{\texttt{Lenstool}}
\newcommand{\ppxf}{\texttt{pPXF}}

\newcommand{\abell}{\text{A383}}
\newcommand{\abellfull}{\text{Abell\,383}}
\newcommand{\ms}{\text{MS\,2137}}
\newcommand{\msfull}{\text{MS\,2137$-$23}}
\newcommand{\mstwo}{\text{MACS J0326}}
\newcommand{\msthree}{\text{MACS J1427}}
\newcommand{\mstwofull}{\text{MACS\,J0326.8-0043}}
\newcommand{\msthreefull}{\text{MACS\,J1427.6$-$2521}}

\newcommand{\rcore}{r_{\mathrm{core}}}
\newcommand{\rcut}{r_{\mathrm{cut}}}
\newcommand{\rcorestar}{{r^*_{\mathrm{core}}}}
\newcommand{\rcutstar}{{r^*_{\mathrm{cut}}}}
\defcitealias{jauzac2019}{J19}
\defcitealias{newman2013lens}{N13}
\defcitealias{zitrin2015}{Z15}

\usepackage{newtxtext,newtxmath}

\title[Kaleidoscope: Lensing in Radial Arc Clusters]{The Kaleidoscope Survey: Strong Gravitational Lensing in Galaxy Clusters with Radial Arcs}

\author[C. Cerny et al.]{
Catherine Cerny,$^{1,2,7}$\thanks{cecerny@umich.edu}
Mathilde Jauzac,$^{1,2,3,4}$
David Lagattuta,$^{1,2}$
Anna Niemiec,$^{5}$
Guillaume Mahler,$^{8}$
\newauthor
Alastair Edge$^{1}$,
Richard Massey$^{1,2}$,
and Joseph Allingham$^{6}$
\\
$^{1}$Centre for Extragalactic Astronomy, Durham University, South Road, Durham DH1 3LE, UK\\
$^{2}$Institute for Computational Cosmology, Durham University, South Road, Durham DH1 3LE, UK\\
$^{3}$Astrophysics Research Centre, University of KwaZulu-Natal, Westville Campus, Durban 4041, South Africa \\
$^{4}$School of Mathematics, Statistics \& Computer Science, University of KwaZulu-Natal, Westville Campus, Durban 4041, South Africa \\
$^{5}$LPNHE, CNRS/IN2P3, Sorbonne Université, Université Paris-Cité, Laboratoire de Physique Nucléaire et de Hautes Énergies, 75005 Paris, France
\\
$^{6}$Physics Department, Ben-Gurion University of the Negev, P.O. Box 653, Be’er-Sheva 84105, Israel \\
$^{7}$Department of Astronomy, University of Michigan 
1085 South University Avenue 
Ann Arbor, MI 48109, USA \\
$^{8}$ STAR Institute, Quartier Agora - All\'ee du six Ao\^ut, 19c B-4000 Li\`ege, Belgium}

\date{Accepted XXX. Received YYY; in original form ZZZ}

\pubyear{2024}

\begin{document}
\label{firstpage}
\pagerange{\pageref{firstpage}--\pageref{lastpage}}
\maketitle

\begin{abstract}

\noindent We measure the dark matter density profiles of six galaxy clusters: \abell, \msfull, \mstwofull, \msthreefull, MACS J0417.5-1154, and MACS J0949.8+1708. Each cluster contains at least one radial arc, a unique physical feature that allows for more precise measurements of the inner mass profile ($R < 50$ kpc) from strong lensing. We present the first strong lensing analysis for \mstwo~and \msthree. We use a combination of \hst~imaging and VLT/MUSE observations from the  ESO Kaleidoscope Clusters Survey, a large `filler' program, to identify and measure redshifts for multiply-imaged systems and obtain the 2-D stellar velocity dispersion for each centrally-located brightest cluster galaxy (BCG). The BCG kinematics are used to subtract the baryonic mass component from the inner mass profile. We find total mass density profiles consistent with previous works using a combination of strong lensing and BCG kinematics. The overall shape of these profiles appears core-like, with an average dark matter slope measurement of $\gamma\sim0.66$. These results demonstrate the ongoing need for the construction of observational models for galaxy clusters, and show how galaxy-scale kinematics can be used to disentangle baryonic and dark matter concentrations in cluster cores. 

\end{abstract}


\section{Introduction} \label{sec:intro}

Galaxy clusters, which are home to the largest concentrations of mass in our Universe, are the ideal cosmic laboratories for studying the properties of baryonic and dark matter. One way the study of these matter distributions can be accomplished is by utilizing the fact that the high mass density of galaxy clusters distorts and magnifies the light emitted by background sources. This unique process, known as gravitational lensing, occurs when a massive object lying along the line of sight between an observer and a background source bends the path of the light traveling from the background source (for reviews of lensing, see e.g. \citealt{massey2010,kneib2011,hoekstra2013,treu2015,kilbinger2015,bartelmann2017}). This distortion can result in the appearance of multiple images of a background source around the massive object. Quantifying the amplification and the positions of the lensed images allows the total mass distribution of the cluster to be mapped with high accuracy (e.g. \citealt{richard2014,jauzac2014,johnson2014,coe2015,caminha2017,williams2018,diego2018,mahler2018,lagattuta2019,sharon2020}), and modeling the effects of lensing allows for a high-resolution measurement of the dark matter (DM) density profile \citep{kneib2004, broadhurst2005, smith2005, limousin2008, newman2009, richard2010, jauzac2016b, jauzac2018a}. As a result, strong gravitational lensing can be an effective tool for studying the distribution of the dark matter mass component in galaxy clusters, although the region that is well constrained by strong lensing is demarcated by the spatial location of lensed images in relation to the center of the cluster ($R\sim200$ kpc). However, lensing's ability to probe the mass distribution of this inner region is unparalleled, which is useful because the precise shape of the inner dark matter mass profile is not yet well understood.

Numerical collisionless cold dark matter (CDM) simulations, which represent our current physical understanding of the Universe, have generally predicted cluster-scale DM distributions that increase toward the center of the cluster following the shape of the Navarro-Frenk-White (NFW) profile \citep{navarro1996}, where the DM density, $\rho_{\mathrm{DM}}$, increases as $\sim r^{-\gamma}$, where the inner slope, $\gamma$, is equal to 1. Subsequent simulations with higher resolution have suggested that this profile is not necessarily universally applicable, as the inner slope may be slightly shallower \citep{merritt2006, navarro2010, gao2012}. Additionally, recent simulations have shown that adding baryons can also move the inner slope away from $\gamma=1$, as seen in the steeper slopes obtained in studies performed using BAHAMAS, Hydrangea/
Cluster-EAGLE, and TNG \citep{maybebahamas,bahe2019,maybefire}. On a smaller scale, \cite{bose2019} conducted an analysis of the APOSTLE (\citealt{fattahi2016}; \citealt{sawala2016}) and AURIGA (\citealt{grand2017}) projects of the inner profile of dwarf galaxies and obtained slope measurements between $\gamma\sim1.3$ and $1.8$. These disparate measurements all point to the need for observational data to serve as a point of comparison. 

Modeling the \color{black}DM\color{black}~density profile of galaxy clusters can be accomplished by placing constraints on the baryonic and DM mass components near the center of the cluster. This is most effectively done by using strong lensing to probe the gravitational potential at the center of the cluster (\citealt{zitrin2012}), and can be further constrained by the inclusion of stellar kinematic measurements of the brightest cluster galaxy (BCG) (\citealt{kelson2002}), which dominates the baryonic mass budget at the center of the cluster. This combination of lensing and kinematics is uniquely powerful because it allows the distribution of dark and baryonic matter in the center of the cluster to be separated, a crucial \color{black}step\color{black}~for measuring the slope of the inner DM profile. Previous studies using this method have \color{black}found\color{black}~shallow slope values; each of these papers used long-exposure, long-slit spectroscopy and Hubble Space Telescope (\hst) imaging for their models, and parameterize the cluster-scale DM halo using the generalized form of the NFW profile. \cite{sand2004} modeled six clusters and found slope values between $0.52-0.57$; \cite{sand2008} remodeled two of these clusters, Abell 383 and MS2137, using updated methods, and found slope values of around 0.45 for both clusters, with some significant statistical uncertainties in the latter model. \cite{newman2013lens} (hereafter N13), which also remodeled these two clusters as a part of a broader sample of seven clusters, measured an average slope of $0.50 \pm 0.13$. The measurements from \citetalias{newman2013lens} were examined in more detail in \cite{he2020}, using the Hydrangea/Cluster-EAGLE hydrodynamic simulation suite to investigate the ability of this combination of lensing and kinematics to accurately recapture the inner slope measurement. The results of this paper suggest that there is a strong degeneracy between the asymptotic generalized NFW slope and the scale radius, $r_{s}$, which means that incorrectly estimating the scale radius, done using weak lensing in \citetalias{newman2013lens}, could lead to much shallower measurements of the inner DM profile. However, using the mean inner slope of the dark matter density profile, rather than the asymptotic generalized NFW slope, was shown to be a better metric of comparison between observational models and simulated clusters. This result indicates that as long as the \color{black}DM\color{black}~density profile in the center clusters is well constrained by some combination of observational measurements, then the inner slope can be measured accurately. 

However, the values for \color{black}the inner DM\color{black}~slope still vary widely\color{black}, though the reason for this discrepancy is not clearly due to methodology, intrinsic cluster mass distributions, or some other factor\color{black}. Recent papers by \cite{sartoris2020} and \cite{biviano2023} have used stellar kinematic measurements of the BCG in combination with cluster member kinematics, obtained from integral-field unit spectroscopy, as a method to constrain the overall mass profile of the cluster. \citeauthor{sartoris2020} found an inner slope measurement of $0.99 \pm 0.04$ for the cluster Abell S1063, and \citeauthor{biviano2023} found an inner slope of $0.7^{+0.2}_{-0.1}$ for the cluster MACS J1206.2-0847. These divergent measurements indicate that the distribution of DM in galaxy clusters \color{black}may be\color{black}~more diverse than suggested by simulations, which in turn demonstrates the need for more observational models to be made to characterize the potential forms of these distributions. 

In this paper, we present the DM inner slope measurements for six different galaxy clusters, using a combination of strong lensing and kinematic measurements of the BCG. We select these clusters from the ESO Kaleidoscope Clusters survey (PID 0104.A-0801; PI A. Edge), a large `filler' program structured around the identification of bright strong-lensing features in galaxy clusters using new integral-field unit spectroscopy from VLT/MUSE. Each of these clusters has a unique physical feature: a `radial arc', or a lensed galaxy \color{black}image\color{black}~located within 5 kpc of the BCG. We note that these radial arcs are preferentially produced in clusters with shallow inner slopes. The proximity of this lensed image to the BCG serves as an additional constraint on the inner \color{black}total\color{black}~density profile in the strong lensing model, which cannot normally measure this region without extrapolating the lens model inward as this region is baryon-dominated. We also choose to use a dual pseudo-isothermal elliptical profile to model the cluster-scale DM halos (\citealt{elisadottir2007}), which avoids the degeneracy between the scale radius and the inner slope measurement pointed out in \cite{he2020}. 

We present new strong lensing and BCG kinematic velocity dispersion measurements for four clusters: \abellfull~(A383 hereafter), \msfull~(MS2137 hereafter), \mstwofull~ (MACS J0326 hereafter), and \msthreefull (MACS J1427 hereafter). We also include in this paper two more clusters that were observed in the Kaleidoscope survey: MACS J0417.5-1154 (MACS J0417 hereafter), and MACS J0949.8+1708 (MACS J0949 hereafter). We utilize previously published strong lensing models for these two clusters since we have no new strong lensing constraints to add to the existing models, and only add the measurement of the stellar velocity dispersion profile of the BCG to obtain the inner DM slope value. We refer the reader to \cite{jauzac2019} and \cite{allingham2023} for more details on the lens models of MACS J0417 and MACS J0949, respectively.  

The paper is organized as follows. The data, observations, and the creation of the VLT/MUSE catalogues used for each cluster are presented in Section~\ref{sec:obs}. Section~\ref{sec:massmodels} discusses the construction of the mass models, and Section~\ref{sec:kinbcg} details the kinematic modeling for the BCG. In Section~\ref{sec:results}, we present our results, and in Section~\ref{sec:disc}, we discuss their implications and describe future work on this problem. We then conclude in Section~\ref{sec:conc}.

We assume a standard \( \Lambda \)CDM cosmology with \( \Omega_M \)=0.3, \(\Omega_{\Lambda} \)=0.7, and \(H_0\) =70 km s\(^{-1}\) Mpc\(^{-1}\). All magnitudes are measured in the AB system unless stated otherwise.


\section{Observations} \label{sec:obs}

\begin{table*}
	\centering
	\caption{Summary of \hst~observations. The name of the cluster is given in the first column, and the R.A. and Decl. are given in degrees (J2000) in the second and third columns. The band is listed in the fourth column, the PID for each band is given in the fifth column, and the P.I. for the observation is given in the sixth column. Finally, the exposure time is given in the seventh column, and the observation date for the exposure is given in the eigth column.}
	\label{tab.hst}
	\begin{tabular}{lccccccc} 
		\hline
		Name & R.A. & Decl. & Band & PID & P.I. & Exp. time [s] & Obs. date\\
		\hline
\abell & 42.0141667 & -3.5291389 & ACS/F435W   & 12065 & Postman & 4250    & 2010-12-28  \\
 &  &  & ACS/F606W   &      &        & 4210    & 2011-01-18  \\
 &  &  & ACS/F814W   &      &        & 8486    & 2010-12-08  \\
 &  &  & WFC3/F105W  &      &        & 3620    & 2011-01-18  \\
 &  &  & WFC3/F125W  &      &        & 3320    & 2011-01-05  \\
 &  &  & WFC3/F140W  &      &        & 2411    & 2011-01-18  \\
 &  &  & WFC3/F160W  &      &        & 5935    & 2010-11-19  \\
\ms & 325.0632083 & -23.6611667 & ACS/F435W   & 12102 & Postman & 4026    & 2011-09-29  \\
 & 325.0916667 & -23.6325000 & ACS/F606W   & 10635 & Ziegler & 17920   & 2006-05-16  \\
 & 325.0632083 & -23.6611667 & ACS/F814W   & 12102 & Postman & 8132    & 2011-08-21  \\
 &  &  & WFC3/F105W  &      &        & 2814    & 2011-09-09  \\
 &  &  & WFC3/F125W  &      &        & 2514    & 2011-09-02  \\
 &  &  & WFC3/F140W  &      &        & 2311    & 2011-09-09  \\
 &  &  & WFC3/F160W  &      &        & 5029    & 2011-08-21  \\
\mstwo & 51.708118 & -0.7310381 & ACS/F814W  & 11103 & Ebeling  & 500 & 2008-11-13  \\
&  &   & ACS/F606W   &      &         & 500    & 2007-11-17  \\	
\msthree & 216.9144704 & -25.3506079 & ACS/F814W   & 12884 & Ebeling & 1440 & 2014-07-21  \\
MACS J0417 & 64.3945486 & -11.9088174 & ACS/F606W & 12009 & von der Linden & 7152 & 2011-01-20 \\
 &  &  & ACS/F435W & 14096 & Coe & 2000 & 2017-06-16 \\
MACS J0949 & 147.4659359 & 17.1195712 & ACS/F606W & 14096 & Coe & 1013 & 2015-11-20 \\
		\hline
	\end{tabular}
\end{table*}

\subsection{Photometry}
\subsubsection{Hubble Space Telescope Imaging}

We utilize imaging from the {\it{Advanced Camera for Surveys}} ({\it{ACS}}) onboard \hst~obtained from the Cluster Lensing And Supernova survey with Hubble (CLASH) survey \citep{postman2012} in the F814W, F6606W, and F435W pass-bands for A383 and MS2137 in order to identify multiple images (see Table~\ref{tab.hst} for details). We also use imaging from the Wide Field Camera 3 (WFC3) in the F105W, F125W, F140W, and F160W pass-bands in order to perform source identification. All imaging was obtained from MAST\footnote{https://archive.stsci.edu/prepds/clash/}. Basic data reduction procedures were applied to all imaging using {\texttt{HSTCAL}} and standard calibration files. {\texttt{Tweakreg}} was used to register individual frames to a common {\it{ACS}} reference image, after which {\texttt{Astrodrizzle}} was used to co-add the frames together.

The observations of \mstwo~and \msthree~are shallower and consist of fewer bands. F606W and F814W images are available for \mstwo~and are sourced from the \hst~SNAP program 11103 (PI: Ebeling). The single F814W image available for \msthree~is taken from archival data made available by \hst~SNAP program 12166 (PI: Ebeling). 

We additionally use the F606W bands from the Reionization Lensing Cluster Survey (RELICS, PID: 14096, PI: Coe) for both MACS J0949 and MACS J0417 to obtain photometric measurements of the BCG. We also include the F435W band from the \hst~programme 12009 (PI: von der Linden) for MACS J0417 as an additional check on our F606W measurements. These images were not used for any other purpose in the course of constructing the lens models.

A summary of the observations, exposure times, and bands used for each cluster are listed in Table~\ref{tab.hst}. Sources for the first three clusters were identified using \textsc{SExtractor} (\citealt{sextractor}) in dual mode on the F814W and F606W pass-band images. Sources for \msthree~were identified using MUSE spectroscopy. Sources for MACS J0949 and MACS J0417 were taken from their existing lens models.

\subsubsection{\hst~Catalogues}

We create new catalogues of multiple image systems from the \hst~imaging for A383, MS2137, MACS J0326, and MACS J1427. We identify arc systems used for lens modeling within the images based on geometry, color and morphology. The properties of the arcs for these four clusters are listed in Tables~\ref{tab.a383arcs}-\ref{tab.ms1427arcs}. For A383, MS2137, and MACS J0326, we build a galaxy catalogue for each cluster using \textsc{SExtractor} in dual mode on the F814W and F606W pass-band images, with threshold parameters {\texttt{DETECT\_THRESH}} = 1.5 and {\texttt{DETECT\_MINAREA}} = 20. Detections with error flags or unreliable magnitude measurements (i.e. {\texttt{MAG\_AUTO=-99}}) were dropped, and detections with a stellarity measurement greater than 0.5 were removed as they are more likely to be stars rather than galaxies. Only objects that appeared in both pass-bands were kept in the final catalogue. Further information on the construction of the galaxy catalogue can be found in Section~\ref{sec:clustergals}. For MACS J1427, we build a galaxy catalogue based on the MUSE detections. For MACS J0417 and MACS J0949, we use the galaxy and arc system catalogues from their existing lens models.

\begin{table}
	\centering
	\caption{Properties of the multiple images that were used as  constraints  in  the  lens  model  of \abell.  R.A. and Decl. are given in degrees (J2000). The redshifts for each image are either the spectroscopic value, where the redshift has no error bars, or the lensing model output value, in which case error bars are provided. The rms for the best fit is measured in the image plane for each family of multiple images. The apparent magnification $\mu$ of each multiple image is also listed.}
	\label{tab.a383arcs}
	\begin{tabular}{lccccccc} 
		\hline
		ID & R.A. & Decl. & $z$ & rms (") & $\mu$ \\
		\hline
1.1   & 42.0128100  & -3.5257360 &  4.63  &   0.87  & 24.9 $\pm$ 5.3  \\
1.2   & 42.0100370 &  -3.5306870 &  4.63  &   0.30  &  >50  \\ 
1.3   & 42.0094730 &  -3.5284480 &  4.63  &   0.67  &  12.9 $\pm$ 4.7 \\
1.4   & 42.0201622 &  -3.5313367 &  4.63  &   0.71  &  4.4 $\pm$ 2.2 \\
2.1   & 42.0100213 &  -3.5312905 &  1.01  &   0.99  &  >50 \\
2.2   & 42.0118119 &  -3.5328386 &  1.01  &   0.95  &  13.8 $\pm$ 1.8 \\
2.3   & 42.0143250 &  -3.5288310 &  1.01  &   1.52  &  1.8 $\pm$ 0.5 \\
3.1   & 42.0124986 &  -3.5352872 &  2.55  &   0.56  &  8.1 $\pm$ 0.7 \\
3.2   & 42.0095063 &  -3.5331842 &  2.55  &   0.39  &  16.6 $\pm$ 4.5  \\
3.3   & 42.0100403 &  -3.5332656 &  2.55  &   0.74  &  2.6 $\pm$ 1.3 \\
3.4   & 42.0159643 &  -3.5351566 &  2.55  &   0.62  & 6.2 $\pm$ 5.6  \\
4.1   & 42.0092467 &  -3.5339770 &  2.55  &   0.09  & 8.2 $\pm$ 3.2  \\
4.2   & 42.0091187 &  -3.5334797 &  2.55  &   0.41  &  16.9 $\pm$ 4.1 \\
4.3   & 42.0117750 & -3.5352866  &  2.55  &   0.42  & 6.7 $\pm$ 2.3  \\
5.1   & 42.0136400 & -3.5263550  &  6.03  &   0.87  & 9.5 $\pm$ 3.4  \\
5.2   & 42.0191904 & -3.5329396  &  6.03  &   0.71  &  4.8 $\pm$ 4.3 \\
6.1   & 42.0177121 & -3.5314173  &  $1.55 \pm 0.28$ & 0.56  &  15.3 $\pm$ 8.7\\
6.2   & 42.0139503 & -3.5332126  &  $1.55 \pm 0.28$ & 0.36  & 15.2 $\pm$ 7.8  \\
6.3   & 42.0088477 & -3.5280946  &  $1.55 \pm 0.28$ & 0.20  &  7.7 $\pm$ 3.2 \\
6.4   & 42.0153782 & -3.5267347  &  $1.55 \pm 0.28$ & 0.40  &  3.1 $\pm$ 1.7 \\
7.1   & 42.0170194 & -3.5239029  &  $4.15 \pm 0.79$ & 0.87  &  18.3 $\pm$ 9.6  \\
7.2   & 42.0148667 & -3.5231278  &  $4.15 \pm 0.79$ & 0.96  &  >50 \\
7.3   & 42.0130417 & -3.5229194  &  $4.15 \pm 0.79$ & 1.05  & 11.0 $\pm$ 3.6  \\
8.1   & 42.0153375 & -3.5235164  &  $1.75 \pm 0.46$ & 0.33  &  >50 \\
8.2   & 42.0141083 & -3.5232670  &  $1.75 \pm 0.46$ & 0.37  &  >50 \\
9.1   & 42.0165440 & -3.5331830  &  $4.27 \pm 1.37$ & 1.12  & >50  \\
9.2   & 42.0171721 & -3.5326837  &  $4.27 \pm 1.37$ & 0.04  &  18.5 $\pm$ 9.6 \\
9.3   & 42.0161051 & -3.5264773  &  $4.27 \pm 1.37$ & 0.72  &  12.6 $\pm$ 4.6 \\
9.4   & 42.0078024 & -3.5279351  &  $4.27 \pm 1.37$ & 0.89  &  5.0 $\pm$ 2.2 \\
		\hline
	\end{tabular}
\end{table}

\begin{table}
	\centering
 	\caption{Properties  of  the multiple images  that  were  used  as  constraints  in  the  lens  model  of \ms. The format of each column is the same as the format for Table~\ref{tab.a383arcs}.}
	\label{tab.ms2137arcs}
	\begin{tabular}{lccccccc} 
		\hline
		ID & R.A. & Decl. & $z$ & rms (") & $\mu$\\
		\hline
1.1   & 325.0653010 & -23.6627183 & 3.086  & 0.24  &  3.8 $\pm$ 1.5  \\
1.2   & 325.0573784 & -23.6552507 & 3.086  & 0.97  &  3.2 $\pm$ 1.8  \\
1.3   & 325.0639173 & -23.6617792 & 3.086  & 0.76  &  1.1 $\pm$ 0.45  \\
2.1   & 325.0627881 & -23.6595561 & 1.19   & 0.27  &  31.7 $\pm$ 8.0  \\
2.2   & 325.0660593 & -23.6669070 & 1.19   & 0.16  &  2.8 $\pm$ 1.1  \\
3.1   & 325.0647182 & -23.6572985 & 1.495  & 0.08  &  7.9 $\pm$ 2.8  \\
3.2   & 325.0623619 & -23.6570179 & 1.495  & 0.77  &  50.2 $\pm$ 7.1  \\
3.3   & 325.0667908 & -23.6653782 & 1.495  & 0.68  &  3.9 $\pm$ 1.1  \\
3.4   & 325.0590207 & -23.6614473 & 1.495  & 0.26  &  3.5 $\pm$ 1.3  \\
4.1   & 325.0617177 & -23.6570091 & 1.495  & 2.07  &  17.3 $\pm$ 2.4  \\
4.2   & 325.0654996 & -23.6574819 & 1.495  & 0.43  &  3.5 $\pm$ 1.9  \\
4.3   & 325.0671724 & -23.6648956 & 1.495  & 0.13  &  4.5 $\pm$ 1.5  \\
4.4   & 325.0671724 & -23.6648956 & 1.495  & 0.41  &  2.7 $\pm$ 1.1  \\
5.1   & 325.0631385 & -23.6593017 & 1.496  & 0.40  &  3.8 $\pm$ 0.84   \\
5.2   & 325.0630167 & -23.6601998 & 1.496  & 0.73  &  0.59 $\pm$ 1.1  \\
5.3   & 325.0649681 & -23.6677483 & 1.496  & 1.43  &  2.6 $\pm$ 1.6  \\
		\hline
	\end{tabular}
\end{table}

\begin{table}
	\centering
  	\caption{Properties  of  the multiple images  that  were  used  as  constraints  in  the  lens  model  of \mstwo. The format of each column is the same as the format for Table~\ref{tab.a383arcs}.}
	\label{tab.ms0326arcs}
	\begin{tabular}{lccccccc} 
		\hline
		ID & R.A. & Decl. & $z$ & rms (") & $\mu$ \\
		\hline
1.1   & 51.70533   & -0.73235   & 3.755  & 1.57  &  9.2 $\pm$ 0.1  \\
1.2   & 51.7140244 & -0.7348069 & 3.755  & 0.96  &  3.2 $\pm$ 2.2  \\
1.3   & 51.7072192 & -0.7311556 & 3.755  & 0.38  &   6.0 $\pm$ 0.16 \\
2.1   & 51.7055216 & -0.7305184 & 1.248  & 0.48  &  22.7 $\pm$ 5.3 \\
2.2   & 51.7057628 & -0.7316436 & 1.248  & 0.56  &  11.7 $\pm$ 3.4 \\
2.3   & 51.7082900 & -0.7303000 & 1.248  & 0.54  &  4.2 $\pm$ 0.6  \\
2.4   & 51.71047   & -0.73444   & 1.248  & 0.93  &  22.6 $\pm$ 4.3  \\
3.1   & 51.7018015 & -0.7310976 & 5.878  & 0.85  &  4.1 $\pm$ 2.3  \\
3.2   & 51.7067455 & -0.7370053 & 5.878  & 0.65  &  6.7 $\pm$ 2.3   \\
3.3   & 51.7048216 & -0.7359669 & 5.878  & 0.29  &  22.0 $\pm$ 10.5  \\
3.4   & 51.7106978 & -0.7288662 & 5.878  & 0.53  & 7.9 $\pm$ 2.3   \\
3.5   & 51.7093690 & -0.7301727 & 5.878  & 0.59  & 11.3 $\pm$ 3.4   \\
		\hline
	\end{tabular}
\end{table}

\begin{table}
	\centering
  	\caption{Properties  of  the multiple images  that  were  used  as  constraints  in  the  lens  model  of \msthree. The format of each column is the same as the format for Table~\ref{tab.a383arcs}.}
	\label{tab.ms1427arcs}
	\begin{tabular}{lccccccc} 
		\hline
		ID & R.A. & Decl. & $z$ & rms (") & $\mu$ \\
		\hline
1.1   & 216.9148671 & 216.9148671 & 0.8836  & 0.24   &  1.9 $\pm$ 1.4  \\
1.2   & 216.9148671 & 216.9148671 & 0.8836  & 0.24   &  11.2 $\pm$ 3.3 \\
1.3*  & 216.9165342 & -25.3486428 & 0.8836  & 0.24   &  3.7 $\pm$ 1.9  \\
2.1   & 216.9106579 & -25.3536281 & 1.23655 & 0.174  &  7.6 $\pm$ 2.8  \\
2.2   & 216.9199276 & -25.3461304 & 1.23655 & 0.174  &  25 $\pm$ 5.1  \\
2.3*   & 216.9131520 & -25.3518332 & 1.23655 & 0.174 &  1.2 $\pm$ 1.1  \\
		\hline
	\end{tabular}
\end{table}

\begin{table*}
	\centering
	\caption{Summary of VLT/MUSE observations used. The name of the cluster is given in the first column, the seeing for the observation is given in the second column, and the airmass for the observation is given in the third column. The P.I. for the observation is given in the fourth column, the exposure time is given in the fifth column, the observation date is given in the sixth column, and the ESO project code is given in the seventh column.}
	\label{tab.muse}
	\begin{tabular}{lcccccc} 
		\hline
		Cluster & Seeing & Airmass & P.I. & Exp. time [s] & Obs. date & ESO Project Code\\
		\hline
A383 & 1.0"   & 1.74 & Edge & 2910    & 2019-11-17 & 0104.A-0801(B) \\
MS2137 & 1.0"   & 1.86 & Edge & 2910    & 2019-06-28 & 0103.A-0777(A)  \\
MACS J0326 & 0.5"   & 1.09-1.11 & Edge & 2910    & 2019-09-21 & 0103.A-0777(A)  \\
MACS J1427 & 1.0"-1.5"   & 1.01-1.06 & Edge & 2910    & 2018-03-14 & 0100.A-0792(A) \\
MACS J0417 & 1.6"  & 1.8 & Edge & 2910 & 2017-12-12 & 0100.A-0792(A) \\
MACS J0949 & 0.71" & 1.4 & Edge & 2910 & 2020-02-20 & 0104.A-0801(A) \\
		\hline
	\end{tabular}
\end{table*}



\subsection{Spectroscopy}
\subsubsection{VLT-MUSE Observations}

The VLT/MUSE observations for each cluster are summarized in Table~\ref{tab.muse}. Each cluster observation consisted of three individual exposures (imaged sequentially) of 970 s each. To minimize the effect of observational systematics, we apply a small dither (0.3 arcsec) between each exposure, and each frame is rotated 90 degrees clockwise relative to the previous frame. The observations were then stacked together to create a single cube with a total exposure of 2910 s. The resulting average seeing and airmass of the stacked cube are reported in Table~\ref{tab.muse}.

Data reduction of the MUSE cubes was performed using the standard procedures of the {\texttt{esorex}} pipeline (\textsc{muse-kit-2.4.1}; \citealt{weilbacher2016}), along with additional calibration and cleaning steps (as described in e.g., \citealt{richard2021} or \citealt{lagattuta2022}). Bias subtraction and flat fielding were performed with basic calibration files using illumination and twilight exposures with dates closest to that of the source exposure. Flux calibration and telluric correction were performed with the standard star taken closest to the date of the source exposure. After an initial reduction process to align individual exposures, we re-run the final calibration step (the "scipost" phase) to improve flux variation between individual IFU slices. This is achieved using an auto-calibration algorithm included in the MUSE reduction pipeline, but we first apply a mask to eliminate flux from bright cluster members and intra-cluster light that bias the measurement. Finally, we apply the \textsc{Zurich Atmospheric Purge} (ZAP; \citealt{soto2016}) to the fully reduced final data cube in order to eliminate strong skyline residuals after sky subtraction.


\subsubsection{VLT-MUSE Catalogues}

\begin{figure} 
    \centering%
        \includegraphics[width=1.0\linewidth]{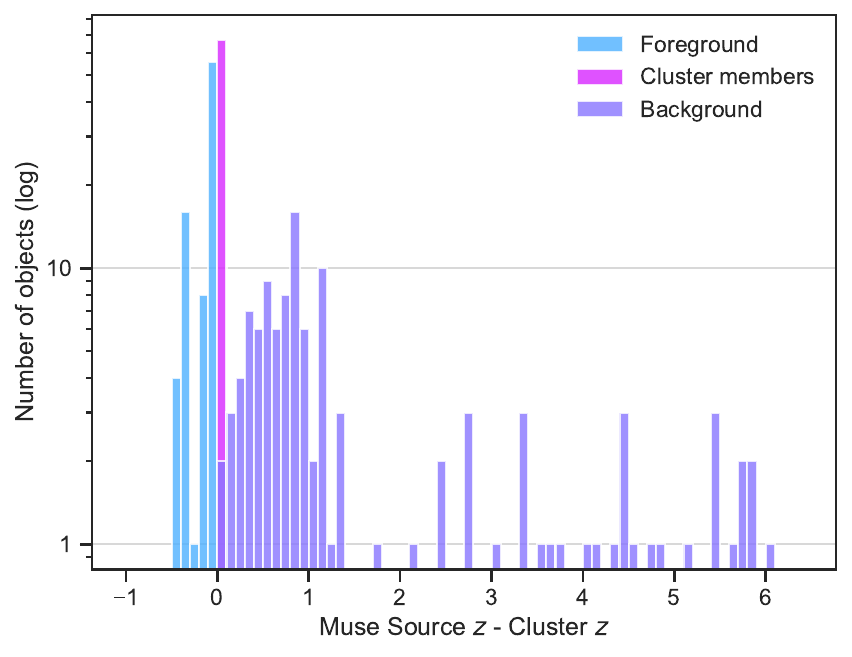}
        \caption{Distribution of MUSE source redshifts for all clusters modeled in this paper: A383, MS2137, \mstwo, and \msthree. Redshifts for MACS J0949 and MACS J0417 are not included as we did not perform source inspection for these clusters. All redshifts are plotted relative to 0, where 0 is equivalent to the cluster redshift and thus denotes all the cluster members identified via MUSE.
        }\label{fig.sourcehist}
\end{figure}
We performed source extraction for four clusters following the procedure detailed in \cite{lagattuta2022}; we briefly describe the procedure here. The clusters we evaluated were A383, MS2137, \mstwo, and \msthree. Spectroscopic redshifts were obtained for objects in the MUSE cubes with a proprietary program called \textsc{Source Inspector}, developed and hosted by CRAL (Centre de Recherche Astrophysique de Lyon). \textsc{SExtractor} \citep{bert96} was first run on the MUSE data using the \texttt{muselet} routine from \texttt{mpdaf} (\citealt{bacon2016}; \citealt{piqueras2017})\footnote{https://mpdaf.readthedocs.io/en/latest/muselet.html}, which identifies sources with flux above the local continuum level by subtracting the average flux measured around a narrow wavelength range from the average flux within that wavelength range. This process creates a pseudo-narrow-band for each wavelength range, or slice. A narrow-band cube was formed by combining all the slices together, and \textsc{SExtractor} was run on each slice to identify emission peaks. Unique detections in the cube had their spectrum extracted after peaks at different wavelength slices that are spatially `close' to each other (i.e. within the same seeing disk) were combined into multi-line objects. Single-line objects, like Ly-$\alpha$ and [OII], were also extracted. The spectra of sources that were identified using bright objects in the existing \hst~imaging were also extracted. A final sky correction was applied by subtracting, from each spectrum, the sum of the 500 nearest blank spaxels (spaxels not associated with any detection in the field) that were located within a 0.4"-4.0" circular annulus centered on the target object.

Each spectrum was then evaluated using \textsc{Source Inspector}, which calculated five possible redshift fits for each spectrum using the tool \texttt{Marz} \citep{hinton2016}. Redshift fits were visually inspected by three individual users, with each user able to select one of the \texttt{Marz} values or manually enter a different value. Each fit was assigned a confidence `rating' between 0-3, where 3 is a confirmed detection (redshift identified from multiple features or one unambiguous feature, such as a Ly-$\alpha$ or [OII] doublet), 2 is a probable detection (several lines that are noisy, which boosts the redshift error, or a single feature that is probably known but could also be something else (i.e. a blended [OII] line that could also be a wide [OIII] line or noisy Ly-$\alpha$), 1 is a possible detection (a best guess, though this is very uncertain), and 0 is no detection (no features, just noise). The selections of the three users were then evaluated against each other, and a complete catalogue was created for each cluster based on the agreement between these selections. Redshifts with a confidence rating of 3 were strongly agreed upon by all three users; redshifts with a confidence rating of 2 were tentatively agreed upon by all three users; and redshifts with a confidence rating of 1 were included as a `best guess'. Confidence 1 redshifts were not included in the final lens models unless they were assigned to lensed galaxies whose positions were supported by the structure of the lens model. A full catalogue for the clusters modeled in this paper (A383, MS2137, \mstwo, and \msthree) can be found in the appendix at the end of this paper in Table~\ref{tab.a383MUSE}, Table~\ref{tab.ms2137MUSE}, Table~\ref{tab.ms0326MUSE}, and Table~\ref{tab.ms1427MUSE}, respectively. Catalogues for MACSJ0949 and MACSJ0417 are not included in this paper as we did not perform source inspections for these clusters; we refer the reader to \citeauthor{jauzac2019} and \citeauthor{allingham2023} for details. Arc systems selected from \hst~photometry were confirmed with these redshift catalogues where possible. \color{black}A histogram of all newly identified MUSE sources is given in Figure~\ref{fig.sourcehist}. The redshift desert in MUSE, which spans between $1.5 \lesssim z \lesssim 2.9 $, indicates an area where spectroscopic identifications of galaxies are difficult to acquire, since no strong nebular emission lines from galaxies at these redshifts fall into the wavelength range covered by MUSE. Identification thus relies on the presence of weaker lines that are still distinct from continuum emission. The presence of the redshift desert is represented by the lack of identified objects in this region, as seen in the figure.\color{black}


\section{Mass Modeling} \label{sec:massmodels}

Strong lensing is a powerful tool for modeling the mass distribution of galaxy clusters, but lensing as a technique cannot differentiate between baryonic and dark matter without ancillary information. This can be partially alleviated by breaking up the overall mass distribution into different `clumps' of matter, such as cluster member galaxies or galaxy-scale perturbers, and modeling them as separate distributions from the main dark matter halo of the cluster. The BCG is a distinct component of any cluster mass model, and in the case of radial arcs, it is particularly important to model it separately because it affects the lensing potential of the cluster more strongly (\citealt{newman2011}). The parameterization of the BCG can be done in a few different ways, but in this paper, we elect to utilize a combination of kinematic measurements from VLT/MUSE and photometry from \HST~imaging to place physical constraints on the distribution of mass in the BCG. This combination of kinematics and lens modeling is a useful technique utilized in many different papers to perform this type of analysis, from \citetalias{newman2013lens} to recent papers like \color{black}\cite{bergamini2023}\color{black}. We describe the parameterization of the cluster mass in this section, and in the following section, we detail the kinematic measurements of the BCG.

\subsection{Lenstool}\label{sec:lenstool}
Strong lens models for each cluster in our sample were constructed using the software \lenstool~ \citep{jul07}, an algorithm that models lensing clusters using a parametric approach. One cluster-scale halo and several smaller substructure halos are combined to create the model, where each halo is treated as a pseudo-isothermal ellipsoidal mass distribution (\color{black}dPIE, or\color{black}~ PIEMD; \citealt{limo05}). The parameters for each halo are: the $x$ and $y$ positions of the center, the ellipticity, $e$, the position angle, $\theta$, the core radius, $\rcore$, the effective velocity dispersion, $\sigma$, and the truncation radius of the cluster halo, $\rcut$. 

Markov Chain Monte Carlo (MCMC) sampling is used to sample the posterior density of the model, which is expressed as a function of the likelihood of the model (as described in \citealt{jul07}). This function is minimized as
\begin{equation*}
\\    
\chi^2_{SL} = \sum_i \chi_i^2 
\end{equation*}
where the sum is performed over the different families of multiple images in the model, and $\chi_i^2$, the chi-square value for each multiply imaged source, is given as 
\begin{equation*}
\\
\chi_i^2 = \sum_{j=1}^{n_i} \frac{\bigl( \theta^j_{\mathrm{obs}}-\theta^j({\mathbf p})\bigr) ^2}{\sigma^2_{ij}}    .
\end{equation*}

\noindent $\theta^j_{\mathrm{obs}}$ is the vector position of the observed multiple image {\it j}, $\theta^j$ is the predicted vector position of the image {\it j}, {\it $n_i$} is the number of images in system {\it i}, $\sigma_{ij}$ is the error of the position of image {\it j}, which is fixed to $0.5"$ for multiple images\color{black}, and $\textbf{p}$ represents each parameter optimized within the model.\color{black}~The model with maximum likelihood thus minimizes the distance between the observed and predicted positions of the multiple images. This distance is referred to as the rms. The optimized model is then used to solve for the best fit set of parameters of each halo.  

\subsection{Cluster Member Galaxies}\label{sec:clustergals}

\begin{figure} 
    \centering
        \includegraphics[width=1.0\linewidth]{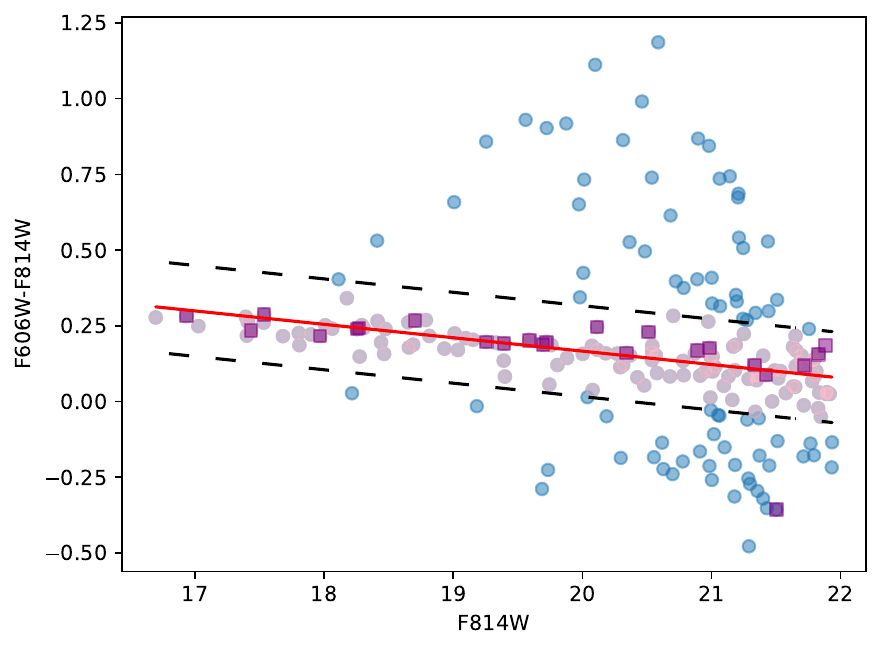}
        \caption{ Example color magnitude diagram for \abell. All sources identified by SExtractor are plotted in blue, and MUSE sources identified around the cluster redshift ($z=0.187$) are over plotted as purple squares. The red line corresponds to the best fit for the red sequence, with a slope of $-0.044\pm0.015$ and an intercept of $1.05\pm0.30$. The boundary  ($\sigma=0.15$) for cluster member selection is marked by the dashed black lines, and the sources identified as cluster members are marked as light purple circles.}
        \label{fig.redseq}
\end{figure}

Cluster member galaxies for all clusters (except \msthree) are selected using the cluster red sequence method \citep{glad2000}, which classifies galaxies as cluster members if they have colors consistent with the red sequence at the cluster redshift. An example red sequence for the cluster A383 is shown in Figure~\ref{fig.redseq}. In the case of \msthree, the archival data available at the necessary resolution to perform red sequence fitting consists of only a single \HST~filter. Red sequence fitting, which requires at minimum two different pass bands, can thus not be performed to isolate cluster members. Instead, we utilize spectroscopic redshifts from MUSE to identify a total of 30 sources at or around the cluster redshift ($0.30 < z < 0.33)$, which we define as the cluster member galaxies for this cluster, with the caveat that this selection can be improved with additional observations. 

For all clusters, we follow the assumption that luminosity traces mass (refer to the discussion in \citealt{harvey2016}) to model each cluster member with galaxy-scale PIEMD halos, where the positional parameters for each halo ($x$, $y$, $e$, $\theta$) are fixed to the properties of their light distribution as measured with \textsc{SExtractor} \citep{bert96}. The remaining PIEMD parameters ($\sigma$, $\rcore$, $\rcut$) are then rescaled to match a reference galaxy with luminosity $L^*$ following the \cite{faberjackson1976} relation:
\begin{equation*}
\\
\begin{cases}
\sigma = \sigma^*\Big(\tfrac{L}{L^*}\Big)^{1/4} \\
\rcore = \rcorestar\Big(\tfrac{L}{L^*}\Big)^{1/2}\\
\rcut = \rcutstar\Big(\tfrac{L}{L^*}\Big)^{1/2}
\end{cases}  \\
\end{equation*}
The mass of each halo is then derived with the following relation:
\begin{equation*}
\\
M = \dfrac{\pi}{G}(\sigma^*)^2{\rcorestar}\Big(\dfrac{L}{L^*}\Big)
\end{equation*}
where $\sigma^*$, $\rcorestar$, and $\rcutstar$~ are the reference velocity dispersion, core radius, and truncation radius, respectively. Previous models have demonstrated that $\rcorestar$~ is small in galaxy-scale halos and has a minimal effect on mass models (e.g. \citealt{covone2006,limousin2007,elisadottir2007}). $\rcorestar$~ is thus fixed to 0.15~kpc for cluster galaxies. The remaining two parameters, velocity dispersion and cut radius, are optimized by the model for a reference galaxy $L^*$ in each cluster. The velocity dispersion is allowed to vary between 27 and $250\mathrm{km s}^{-1}$, and the cut radius between 3 and 50 kpc. The cut radius is constrained to an upper limit in order to account for tidal stripping of galactic dark matter halos \citep{limousin2007a,limousin2009,natarajan2009,wetzel2010,niemiec2019}. We differentiate between models using $\chi^2$ and rms statistics, where a low rms generally indicates a better model.

\begin{figure*} 
    \begin{minipage}{0.70\textwidth}
        \includegraphics[width=0.5\linewidth,height=0.5\linewidth]{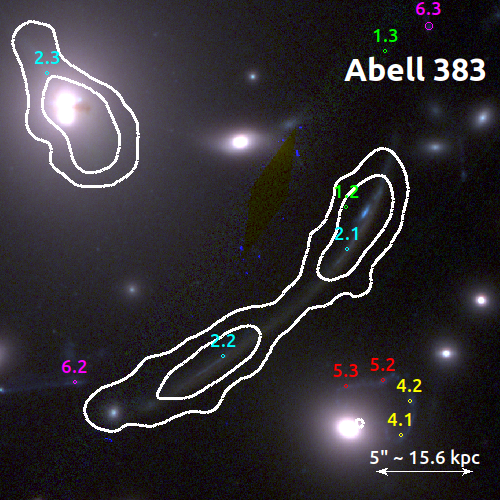}
        \includegraphics[width=0.5\linewidth,height=0.5\linewidth]{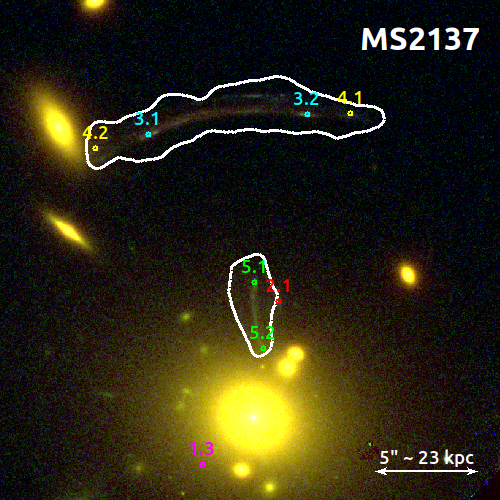}
        \includegraphics[width=0.5\linewidth,height=0.5\linewidth]{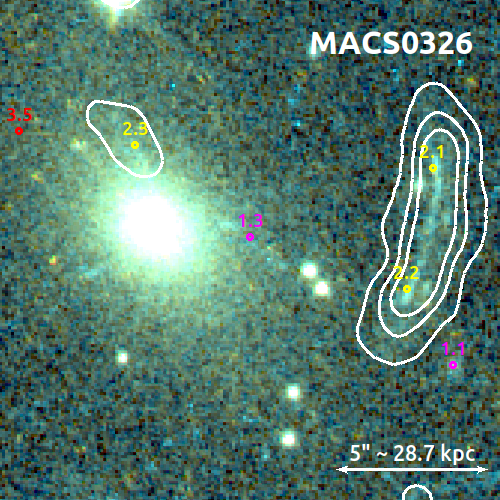}
        \includegraphics[width=0.5\linewidth,height=0.5\linewidth]{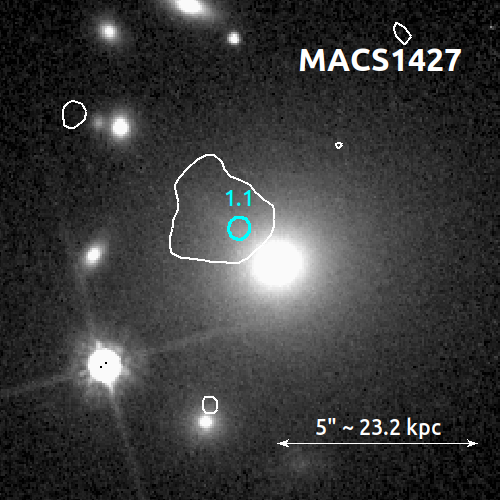}
        \includegraphics[width=0.5\linewidth,height=0.5\linewidth]{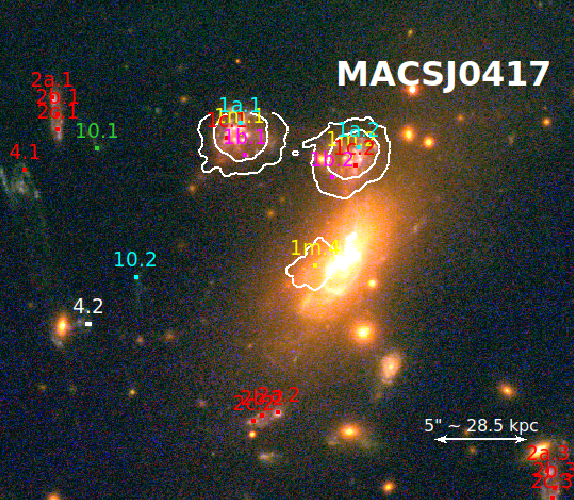}
         \includegraphics[width=0.5\linewidth,height=0.5\linewidth]{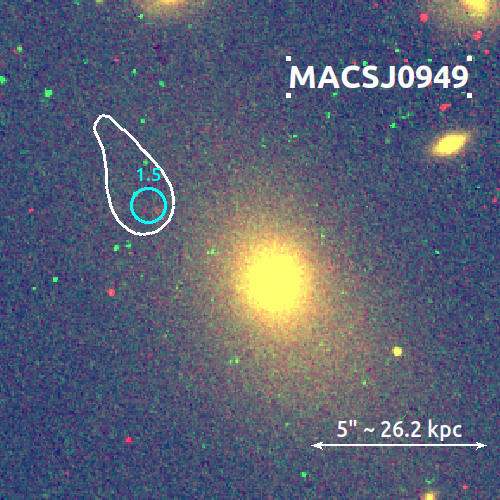}
        \caption{Snapshots of the radial arcs in each cluster. All images are oriented North-up, East-left. Each individual arc in a system is labeled as a pair of numbers, i.e. 1.1. In the case where an arc is labeled, for example, 1a.1, the letter corresponds to a sub-division of the same lensed galaxy 1. Each arc 'family' is marked in a different color for clarity. {\it Top Left}: \abell~image of the BCG with MUSE detection contours for the radial arc and its nearby tangential counterpart at 7495 \AA~overplotted in white. Multiple image systems in this region are also shown. {\it Top Right}: \ms~image of the BCG with MUSE detection contours of the radial arcs and the nearby tangential arc at 9300 \AA~overplotted in white. Multiple image systems in this region are also shown. {\it Middle Left}: \mstwo~image of the BCG with MUSE detection contours of the radial arc at 8376 \AA~overplotted in white. Multiple image systems in this region are also shown. {\it Middle Right}: \msthree~image of the BCG with MUSE detection contours of the radial arc at 7020 \AA~overplotted in white. Multiple image systems in this region are also shown. {\it Bottom Left}: MACSJ0417 image of the BCG with MUSE detection contours of the radial arc and its nearby counterpart images at 6975 \AA~overplotted in white. Multiple image systems in this region are also shown. {\it Bottom Right}: MACSJ0949 image of the BCG with MUSE detection contours of the radial arc at 7160 \AA~overplotted in white. Multiple image systems in this region are also shown.
        }\label{fig.giantarcs}
    \end{minipage}%
\end{figure*}

\subsection{Multiple Images}
\color{black}Each\color{black}~ model is constrained using the positions of multiply imaged sources, or arcs.
Arcs are identified through a combination of MUSE spectroscopy, visual identification of sources with matching color and morphology, and, where relevant, referencing previous models constructed for these clusters. The positions for each source with spectroscopy were fixed to the MUSE detections, which was especially relevant for each of the radial arcs, as their exact substructure and shape is obscured by the BCG light \color{black}and can only be disentangled using MUSE.  The properties of the multiple image systems for each cluster modeled in this paper are described in Table~\ref{tab.a383arcs}, Table~\ref{tab.ms2137arcs}, Table~\ref{tab.ms0326arcs}, and Table~\ref{tab.ms1427arcs}.\color{black} 

\color{black}\subsection{Lens Models} 
A summary of the MCMC fit values for each cluster can be found in Table~\ref{tab.modelstats}.\color{black}
\subsubsection{\abell}

\abellfull~($z=0.187$) was first modeled using \lenstool~in \cite{sand2004}. It was later remodeled in \cite{sand2008} in response to criticism leveled against the initial modeling method's assumptions ignoring cluster substructure, and using spherically symmetric mass distributions; the 2008 model used a full 2D strong lensing model to avoid making these assumptions. The cluster was then observed by CLASH (\citealt{clash}), and the discovery of a multiply-imaged system at $z=6.027$ was reported in \cite{richard2011}, while a mass model was presented in \cite{newman2011} and \cite{newman2013lens}. \cite{zitrin2011} then identified four new multiple image systems and presented an updated model, which was later refined with additional spectroscopy for one system by \cite{zitrin2015}.

A total of nine systems of multiple images are used to constrain the \color{black}current\color{black}~ lens model. The properties of the systems are presented in Table~\ref{tab.a383arcs}. Systems 1, 2, and 9 are all fixed to the spectroscopic redshift measured and optimized from the MUSE cubes. The redshifts of the remaining systems were solved for by the lens model. Systems 2-9 are used in previous lens models of this cluster \citep{sand2008, zitrin2011, newman2011, richard2011, newman2013lens, zitrin2015}, while system 1 is a new identification confirmed by MUSE spectroscopy. The spectroscopic redshifts for the multiple images are listed in Table~\ref{tab.a383MUSE}.

\subsubsection{MS2137$-$223}
\msfull~($z=0.313$) was first modeled with \lenstool~ in \cite{sand2004} and then updated in \cite{sand2008} using a full 2D strong lensing model. \cite{donnarumma2009} presented a mass profile for the cluster based on strong gravitational lensing and {\it{Chandra X-ray Observatory}} imaging. \cite{newman2013density} later created a mass model with similar imaging and new Keck spectroscopy for seven multiply-imaged sources, and extended stellar velocity dispersions for the BCG. The lack of multiple images in these models caused them to all have slightly different results for the mass profiles. CLASH observations (\citealt{clash}) were later carried out on the cluster, and a new model with an additional multiply-imaged system with a confirmed spectroscopic redshift was published in \cite{zitrin2015}. A KMOS study has also recently been performed that includes several galaxies in \ms~\citep{2020MNRAS.496..649T}.

We use a total of four systems of multiple images to optimize this lens model. The properties of the systems are presented in Table~\ref{tab.ms2137arcs}. All systems have redshifts from MUSE observations. System 1 was first identified in \cite{zitrin2015}, but only with photometric information. We confirm its redshift here with MUSE spectroscopy. The remaining systems are referenced in previous works \citep{sand2008, donnarumma2009, newman2013lens, zitrin2015}. We note that while system 2 is not a new detection, this is the first time this arc has been included in a parametric lens model. We are able to include it because it has a confirmed redshift measurement from MUSE spectroscopy. Systems 3 and 4 are parts of the same giant radial arc, and are adopted here as separate constraints in order to refine our lens model. 

\begin{table*}
	\centering
	\caption{Values for the BCG of each cluster extracted from the photometric fitting process. The first column lists the cluster, and the second column lists the filter used for the photometric fit of the BCG. The third column lists the $b/a$ value, the fourth column lists the position angle $\theta$, the fifth column lists the magnitude of the BCG in the listed filter, and the sixth and seventh columns list the $r_\mathrm{core}$ and $r_\mathrm{cut}$ fit values. The errors for $r_\mathrm{core}$ are extremely small and are thus not listed. }
	\label{tab.bcgphot}
	\begin{tabular}{lcccccc} 
		\hline
		Cluster & Filter & $b/a$ & $\theta$ & Magnitude & $r_{\text{core}}$ [kpc] & $r_{\text{cut}}$ [kpc] \\
		\hline
A383        & ACS/F606W  & 0.87   & 81.77 & 18.82 & 3.32    & 23.8 $\pm~ 5.6$  \\
MS2137      & ACS/F625W  & 0.86   & 73.56 & 18.40 & 0.76    & 66.2 $\pm~ 12.3$   \\
\mstwo & ACS/F606W  & 0.55   & 46.66 & 19.56 & 1.74  & 77.9  $\pm~ 6.4$  \\
\msthree & ACS/F814W  & 0.90   & -2.85 & 17.88 & 0.23    & 7.0  $\pm~ 0.31$   \\
MACS J0949   & ACS/F606W  & 0.52   & 31.12 & 19.96 & 2.81    & 70.6 $\pm~ 4.2$ \\
MACS J0417    & ACS/F606W  & 0.33   & -34.10 & 19.58 & 6.85  & 101.4 $\pm~29$  \\
		\hline
	\end{tabular}
\end{table*}

\subsubsection{MACSJ 0326.8-0043}
\mstwo~($z=0.447$) was first discovered by the MAssive Cluster Survey SNAPshot programs (MACS, PI: Ebeling, \citealt{ebeling2001},  \citealt{macs2012}), which amassed a sample of the most X-ray luminous galaxy clusters based on X-ray sources detected by the Röntgen Satellit (ROSAT) All-Sky Survey \citep{voges1999}. These programs were designed to carry out \hst~observations of very X-ray luminous sources to obtain a statistically robust sample of massive distant clusters of galaxies. This\color{black} is the first published strong lensing mass\color{black}~ model for this cluster.

Three systems are used in this lens model, with all systems having at least one image redshift from MUSE. The properties of the arc systems are presented in Table~\ref{tab.ms0326arcs}. All three arcs are new identifications as this cluster has not previously been modeled. System 2 is a clear tangential arc located North-East of the BCG, and has been split into a northern and southern part for the purposes of this model. These arcs, 2.1 and 2.2, are MUSE detections, as is 2.3, which is the Northern radial arc shown in Figure~\ref{fig.giantarcs}. Image 2.4 is predicted by the model and is located on a similarly-colored source. Systems 1 and 3 are Lyman-$\alpha$ emitters. Images 3.1, 3.2, and 3.3 are MUSE \color{black}identifications\color{black}, while 3.4 and 3.5 are predicted multiple images from the model. Image 1.2 is a MUSE identification, while images 1.3 and 1.4 are fixed to locations predicted by the model and are confirmed by the presence of similarly-colored sources in the image. Image 1.4 is of particular interest, as it hints at another radial arc next to the BCG. Figure~\ref{fig.giantarcs} seems to indicate the presence of another source in this area, but the shallowness of the \hst~image makes it difficult to confirm. Deeper spectroscopic observations would be able to confirm the legitimacy of this arc. 

\subsubsection{MACSJ 1427.6$-$22521}
\msthree~($z=0.318$) was first observed in the MACS program (\citealt{ebeling2001}). This is the first published \color{black}strong lensing\color{black}~ mass model for this cluster.

Two multiple image systems are used to construct the lens model; the first is the radial arc located near the BCG (1.1), with a spectroscopic redshift of $z=0.884$, while the second is another \color{black}source with a spectroscopic redshift of\color{black}~ $z=1.237$ (2.1). The properties of these systems are presented in Table~\ref{tab.ms1427arcs}. \color{black}The multiple image systems used to construct the lens model for this cluster are based primarily on the identification of these two sources in the MUSE cube, as visual confirmation of their color and morphology is not possible within the single available band of \hst\ imaging. The lens model predicts additional counter-images for these systems, but there is no significant emission present in the MUSE cube at these locations that would allow for the verification of these counter-images, and inspection of the \hst\ imaging also does not confirm or rule out their existence.\color{black}~ The predicted positions \color{black}of the counter-images\color{black}~ are listed in Table~\ref{tab.ms1427arcs}, but should be treated as potential candidates rather than a confirmed identification \color{black}of a lensed arc\color{black}.

\subsection{Additional Clusters with Radial Arcs}
The clusters presented thus far are not the only clusters in the larger Kaleidoscope sample of MUSE cubes that have radial arcs. In the interest of completeness, we introduce the two remaining clusters in the sample \color{black}with radial arcs\color{black}. These clusters have both been modeled using \lenstool~\color{black}outside this paper\color{black}~ following the methods already described in \color{black}Section~\ref{sec:lenstool}\color{black}, which allows us to directly incorporate them into our broader analysis of the density profiles of galaxy clusters with radial arcs. These two clusters are MACS J0417 and MACS J0949. Both clusters were first observed by MACS (\citealt{ebeling2001}; \citealt{macs2012}). 

MACS J0417 was modeled in \citealt{mahler2019} as a part of the Reionization Lensing Cluster Survey (RELICS; \citealt{coe2019}), as well as in \citealt{jauzac2019} (J19), which included MUSE spectroscopy. We utilize the fiducial model from \citetalias{jauzac2019} in our analysis. The cluster is at a redshift of $z=0.443$, and is constrained with a total of 17 lensed systems, three of which have spectroscopic redshifts confirmed by MUSE. System 1, a quadruply-\color{black}imaged\color{black}~ galaxy, is of particular interest, as it includes a radial arc whose position is identified from emission lines detected in MUSE, as it is otherwise obscured by the BCG in the available \hst~imaging. Unlike the other five clusters in our sample, MACS J0417 has a fairly elongated cluster core, which can be observed both visually by the separation in projection of the second and third brightest cluster galaxies from the BCG, as well as in the X-ray analysis (see \citetalias{jauzac2019}), which show extended emission along the SE-NW axis. These factors indicate that the cluster is likely the result of a recent merger, possibly oriented along the line of sight, which makes it the most dynamically complex cluster of our sample.

Finally, we include MACS J0949 in our sample, using the strong lens model recently published in \cite{allingham2023}. The cluster is at a redshift of $z=0.383$, and X-ray data suggests that it is a post-merger in the process of relaxing. The strong lens model is constrained with a total of 6 lensed systems, two of which have spectroscopic redshifts confirmed by MUSE. Similarly to the other clusters in this paper, its radial arc is also detected by MUSE.

\section{BCG Kinematics}\label{sec:kinbcg}

We introduce kinematic constraints into the total density profile of each cluster in order to more closely examine the shape of the  mass distribution in the inner~10 kpc of the cluster. Following the methods used in \cite{sand2004} and \cite{newman2013density}, we include the velocity dispersion of the BCG as an additional constraint for our \lenstool~model. We incorporate this constraint by measuring observable properties of the BCG through a combination of photometry and spectroscopy. We model the BCG as a separate dPIE mass halo within \lenstool, with its parameters fixed to the values obtained from these measurements. 
The following subsections discuss the measurements of the observable properties of the BCG that we use to fix the parameters of the BCG mass halo, as well as to constrain the lens model using our kinematic measurements. We discuss the combination of kinematics and lensing more thoroughly in Section~\ref{sec.kinlensing}. 

\subsection{BCG Photometry} \label{sec.bcgphot}

\begin{figure} 
    \centering
        \includegraphics[width=1.0\linewidth]{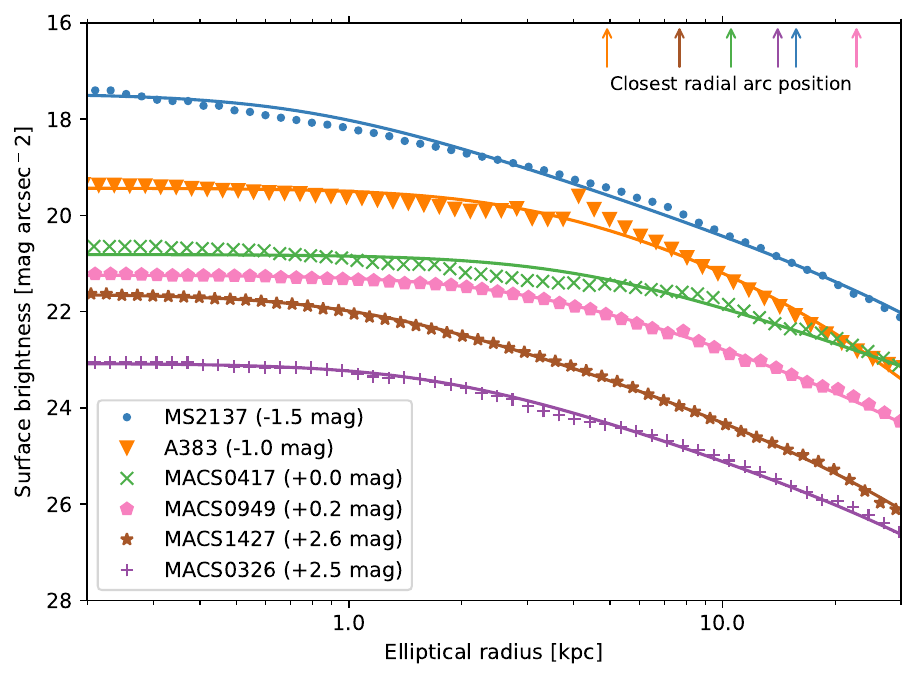}
        \caption{ Surface brightness profiles of the BCGs for all six clusters, measured using the filters listed in Table~\ref{tab.bcgphot}. The dPIE fit for each profile is plotted as the solid line of the same color as each cluster. The magnitudes for each cluster are offset by the listed values to provide visual clarity. \color{black}Color-coded arrows are plotted along the top to show the approximate distance from the BCG to the barycentre of the radial arc in each cluster.\color{black}}
        \label{fig.sbprof}
\end{figure}

We measure the surface brightness profile of the BCG in each cluster using \hst~imaging, where preference was given to the {\it{ACS}}/F606W filter to more closely align the photometric fit with the wavelengths used in the kinematic fit since these BCGs are largely red and dead. This band was used for A383, \mstwo, MACS J0949, and MACS J0417, while the {\it{ACS}}/F814W filter is used for \msthree. However, in MS2137, where the BCG \color{black}is near the chip gap\color{black}, the F625W filter was used. In \msthree, where this band is not available, the F814W filter was used instead. Surrounding objects in the field are masked out using a \textsc{SExtractor} segmentation map, and the PSF is modeled using a field star and a selection of surrounding sky. We model the BCG light using the 2 dimensional dPIE equation, adopting the specific form used in \lenstool~(\citealt{limousin2005}):

\begin{equation*}\label{eq:dpie} 
\\
\Sigma_{\mathrm{dPIE}}(R) = \frac{\sigma_0^2r_{\mathrm{cut}}}{2G(r_{\mathrm{cut}}-r_{\mathrm{core}})} \left( \frac{1}{\sqrt{r_{\mathrm{core}}^2+R^2}} - \frac{1}{\sqrt{r_{\mathrm{cut}}^2+R^2}} \right),
\end{equation*}

\noindent where $r_{\mathrm{core}}$ is the core radius, $r_{\mathrm{cut}}$ is the cut radius, and $\sigma_0$ is the central velocity dispersion of the BCG. The half-light radius, $r_h$, is related to the cut radius as $r_h \approx r_{\mathrm{cut}}$, and in the limit $r_{\mathrm{core}}/r_{\mathrm{cut}} \ll 1$, the projected effective radius is $R_e \approx (3/4)r_{\mathrm{cut}}$. 

We fit this equation by using the light profile extracted from isophotal measurements of the BCG. We first fix the geometric parameters of ellipticity, position angle, and center coordinates of the central isophote to the values we obtain from fitting a de Vaucouleurs $R^{1/4}$ profile to the 2D data using {\texttt{GALFIT}} (\citealt{peng2002,peng2010}). We then use the Python Astropy Elliptical Isophote Analysis routine to find the best-fit model for the 2D data out to about 20 kpc, where the light from other objects in the field begins to dominate the light from the BCG, and run the fit in the bilinear area integration mode. The 1D surface brightness profile is then fit to the dPIE equation, with only $\sigma_0$ allowed to vary during the fit. 

Finally, since we use the mass of the BCG as a constraint in our modeling, we require a stellar mass-to-light ratio to transform the dPIE fit to the light into the mass of the BCG. We obtain this value by performing a SED fit of the combined photometric (see Table~\ref{tab.hst}) and spectroscopic (see Table~\ref{tab.muse}) data using \texttt{ppxf} (\citealt{cappellari2017}). We use the FSPS models library (\citealt{conroy2009}, \citealt{conroy2010}) generated in \cite{cappellari2023} as the basis for the fit, which was created using a Saltpeter IMF with a mass range between 0.08 and 100 \(M_\odot\). These models do not explicitly include the effect of gas or dust. The spectra for these SPS models were created using the MILES stellar library (\citealt{sanchez2006}, \citealt{falcon2011}). We use the library of these spectra to fit the stellar mass-to-light ratio of the BCGs in each cluster.

The resulting parameters, surface brightness profiles, and dPIE fits are presented in Table~\ref{tab.bcgphot} and Figure~\ref{fig.sbprof}. This fit allows us to easily create a separate mass halo for each BCG whose properties are fixed to these photometric values. We specifically fix  the ellipticity, the position angle, and the values of $\rcore$ and $\rcut$. The only free parameter is then the velocity dispersion, $\sigma_0$, which serves as a constraint on the mass profile in our model.

\subsection{Velocity Dispersion Profile}

\begin{table}
	\centering
	\caption{Values for the velocity dispersion profile for the BCG of each cluster. The quoted $\sigma$ values are measured at the midpoint of each bin. The first column lists the cluster, the second column lists the total extent of each bin in arcsec, the third column lists the total extent of each bin in kpc, and the fourth column lists the measured velocity dispersion in that bin.}
	\label{tab.bcgsigma}
	\begin{tabular}{lcccc} 
		\hline
		Cluster & Bin [arcsec] & Bin [kpc] & $\sigma$ [km/s] \\
		\hline
A383        & 0.0-0.26   & 0.0-0.81 & 285 $\pm$ 14 \\
            & 0.46-0.66  & 1.44-2.07 & 329  $\pm$ 12 \\
            & 0.86-1.06  & 2.69-3.32 & 263  $\pm$ 10 \\
            & 1.26-1.46  & 3.94-4.57 & 272  $\pm$ 10 \\
            & 1.66-1.86  & 5.19-5.82 & 295  $\pm$ 13 \\
            & 2.06-2.26  & 6.45-7.07 & 318  $\pm$ 15 \\
            & 2.46-2.66  & 7.70-8.32 & 296  $\pm$ 16 \\
            & 2.86-3.06  & 8.95-9.57 & 322  $\pm$ 17 \\
            & 3.30-3.66  & 10.33-11.45 & 316  $\pm$ 20 \\
            & 4.46-5.06  & 13.96-15.83 & 341  $\pm$ 24 \\
            & 5.86-7.26  & 18.34-22.72 & 409  $\pm$ 28 \\
            \hline
MS2137      & 0.0-0.50   & 0.0-2.30 & 337 $\pm$ 13  \\
            & 0.50-1.0   & 2.30-4.60 & 350 $\pm$ 9 \\
            & 1.0-1.5    & 4.60-6.90 & 334 $\pm$ 8 \\
            & 1.5-2.0    & 6.90-9.20 & 374 $\pm$ 10 \\
            & 2.0-2.5    & 9.20-11.50 & 378 $\pm$ 13 \\
            & 3.0-3.5    & 13.80-16.09 & 404 $\pm$ 19 \\
            \hline
\mstwo    & 0.43-0.73  & 2.47-4.19 & 238 $\pm$ 15    \\
            & 0.73-1.08  & 4.19-6.20 & 276 $\pm$ 15 \\
            & 1.08-1.66  & 6.20-9.52 & 291 $\pm$ 17 \\
            \hline
\msthree    & 0.0-0.38   & 0.0-1.76 & 320 $\pm$ 9   \\
            & 0.38-0.75  & 1.76-3.48 & 312 $\pm$ 9 \\
            & 0.75-1.12  & 3.48-5.19 & 297 $\pm$ 10 \\
            & 1.12-1.5   & 5.19-6.95 & 301 $\pm$ 11 \\
            & 1.5-1.88   & 6.95-8.71 & 326 $\pm$ 16 \\
            & 1.88-2.25  & 8.71-10.43 & 344 $\pm$ 21 \\
            & 2.25-2.62  & 10.43-12.14 & 407 $\pm$ 31 \\
            \hline
MACSJ0949   & 0.0-0.71   & 0.0-3.71 & 300 $\pm$ 12 \\
            & 0.71-1.42  & 3.71-7.43 & 292 $\pm$ 10 \\
            & 1.42-2.13  & 7.43-11.14 & 282 $\pm$ 12 \\
            & 2.13-3.18  & 11.14-16.63 & 323 $\pm$ 24 \\
            \hline
MACSJ0417   & 0.0-0.85   & 0.0-4.85 & 380 $\pm$ 13  \\
            & 0.85-1.7   & 4.85-9.70 & 388 $\pm$ 13 \\
            & 1.7-2.55   & 9.70-14.56 & 399 $\pm$ 15 \\
            & 2.55-3.4   & 14.56-19.40 & 429 $\pm$ 22 \\
		\hline
	\end{tabular}
\end{table}

\begin{figure*} 
    \centering
        \includegraphics[width=1.0\linewidth]{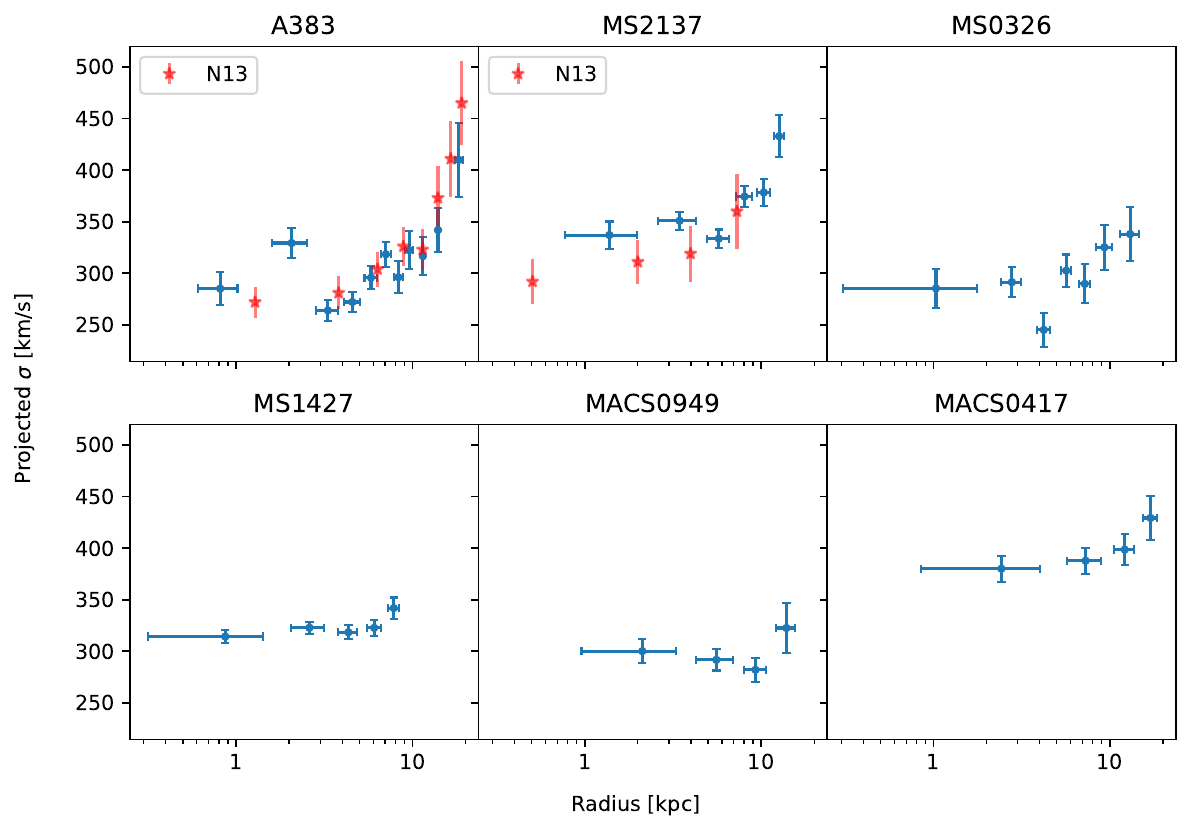}
        \caption{BCG stellar velocity dispersion profiles for all six clusters. The error in the velocity dispersion is denoted by the errorbars along the Y axis, while the errorbars on the X axis signify the width of the bin used to make the measurement. The datapoints are centered at the middle of each bin, and each point corresponds to the reported $\sigma$ value in Table~\ref{tab.bcgsigma}. For the clusters A383 and MS2137, the values reported in \citetalias{newman2013lens} are over-plotted as red stars for reference. } 
        \label{fig.sigmaprof}
\end{figure*}

We utilize BCG kinematics in tandem with lensing to break down the total mass distribution of the cluster into more clearly defined baryonic and dark matter segments. Modeling the BCG mass using stellar dynamics allows us to place limits on the contribution of the BCG to the \color{black}total\color{black}~cluster density profile in the very inner portion of the cluster (<10 kpc), which in turn allows us to more precisely examine the behavior of dark matter in this region. 

We elect to model the kinematics of the BCG by creating a profile of the velocity dispersion in different bins, stepping out from the center of the BCG until the S/N ratio dips below a cutoff threshold of 30, which is equivalent to a distance of about $\sim$20 kpc for each cluster. \color{black}We note that this profile increases with radius because the measurement begins to sample the cluster potential at larger radii; thus only the lower values of $\sigma$ characterize the BCG itself.\color{black}~ We utilize the MUSE observations from the Kaleidoscope survey to make these measurements of the velocity dispersion. As these observations are all short exposures, we choose to bin the cubes using concentric circular annuli, which maximizes the S/N ratio in each bin  \color{black}and is applied to the BCGs in each cluster under our guiding assumption that each BCG can be described by a spherical mass distribution\color{black}~(see Section~\ref{sec:anisotropy} for more details). The width of each annulus is determined by the S/N ratio, where we require each bin to be large enough to obtain a total S/N ratio greater than 30. We fit each spectrum over a rest-frame wavelength of $4860$-$7160 ~{\mbox{\normalfont\AA}}$, which includes the regions where MUSE sensitivity is strongest and excludes regions where sky line residuals are very large.

Nearby bright stars and galaxies are masked out before the fit is performed for each BCG, and bins where these objects interfere strongly with the BCG light are excluded. We choose to mask out surrounding galaxies despite their potential contribution to the velocity dispersion profile because we do not know where they are located in the 3D space of the cluster. Additionally, failing to mask these objects distorts the velocity dispersion toward non-physical values. This is particularly relevant for \msthree, which has a bright foreground star situated within 5" of the BCG. While masking this star allows for an effective measurement of the velocity dispersion, we are unable to continue fitting past about 10~kpc due to the contamination from the flux of this star and the surrounding cluster galaxies.

The \ppxf~fit for each cluster relies on the use of additive and multiplicative polynomials in order to match the template spectra to the observed VLT/MUSE spectrum. These polynomials are adjusted until the fit in the highest S/N bin of each cluster no longer improves. One first-order multiplicative polynomial is used for every cluster except MS2137, which uses a third-order polynomial, and a fifth-order additive polynomial is used for every cluster.

We are able to compare our measurements for the velocity dispersion directly to the results found in \cite{newman2013lens}. Figure~\ref{fig.sigmaprof} shows the velocity dispersion profiles for all six clusters analyzed in this paper, and additionally shows how our results compare to those derived in \citetalias{newman2013lens} for the clusters A383 and MS2137. The \citetalias{newman2013lens} results are based on a Keck/LRIS 23.7 ks exposure with 0.8" seeing for A383, and a Keck/ESI 6.7 ks exposure with 0.7" seeing for MS2137. Relative to the MUSE data, the archival \color{black}Keck\color{black}~ data have a 10x and 3x longer exposure time for A383 and MS2137. Measured seeing conditions were on average 0.2" narrower. However, the advantage of MUSE rests in the ability of IFU observations to capture information about the entire structure of the BCG. In other words, area recovers depth. This is well supported by our measured profiles, because while there is some variation in the innermost bins, the shape of the profile for A383 and MS2137 is generally well reproduced by our measurements. This suggests that our characterization of the BCG profile in the MUSE data is consistent with archival measurements, despite the difference in observing conditions. 

\color{black}
\subsection{Velocity Anisotropy}\label{sec:anisotropy}

The dynamical analysis performed in this paper is accomplished under one critical assumption: that the BCGs in this work can be modeled with the spherical Jeans equation. This simplification relies on the assumption that \color{black} each BCG has an isotropic velocity dispersion and can be described by using spherically symmetric geometry. \color{black}
We choose to rely on this assumption in the creation of the combined kinematic-strong lensing model because BCGs generally lack significant \color{black}anisotropy\color{black}~  (see \cite{kronawitter2000}, which studied 21 giant ellipticals and found that the velocity anisotropy was less than $\beta <0.3$ for the majority of the sample, and \cite{santucci2023}, which measured a similar value for central ellipticals). Choosing to treat the BCGs in this way significantly speeds up the computational time required for the optimization procedure in \lenstool (by a factor of at least 10). However, it is worth considering whether this assumption would have an impact on our mass modeling procedure, as MACS J0417 does have a velocity gradient reaching across the major axis of the BCG, which is likely due to the actively evolving dynamical state of the ongoing merger happening within the cluster.  

To test the validity of this assumption, we employ two different methods to evaluate what effect \color{black}anisotropy\color{black}~would have on our lens models. First, we use the form of the anisotropic Jeans equation and fix the anistropy parameter $\beta$ to a small, constant value ($\beta = 0.3$) to examine how this affects our results. Second, for MACS J0417, we employ the more robust external modeling program JamPy (\citealt{Cappellari2020}) in conjunction with MgeFit (\citealt{Cappellari2002}) to examine the specific effect that the presence of velocity anistropy in the BCG might have on the measurement of the cluster mass profile. Introducing a constant anisotropy value shifts the value of the derived velocity dispersion by roughly $\pm 5$ km/s in all cases. The more robust examination of anisotropy in MACS J0417 results in around the same changes. This small shift is not significant for the measurements presented in this paper, as the deviation is encapsulated in the error budget intrinsic to the models. In future work, it would be worth reexamining this question in the context of mass models that are created using multi-wavelength analysis, such as models that specifically constrain components like the intra-cluster medium or intra-cluster light by incorporating X-ray imaging into the model.

\color{black}

\section{Results}\label{sec:results}

\subsection{Mass Models}

\subsubsection{\abellfull}
\begin{figure*} 
   \centering%
    \begin{minipage}{1.0\textwidth}
        \includegraphics[width=0.5\linewidth,height=0.5\linewidth]{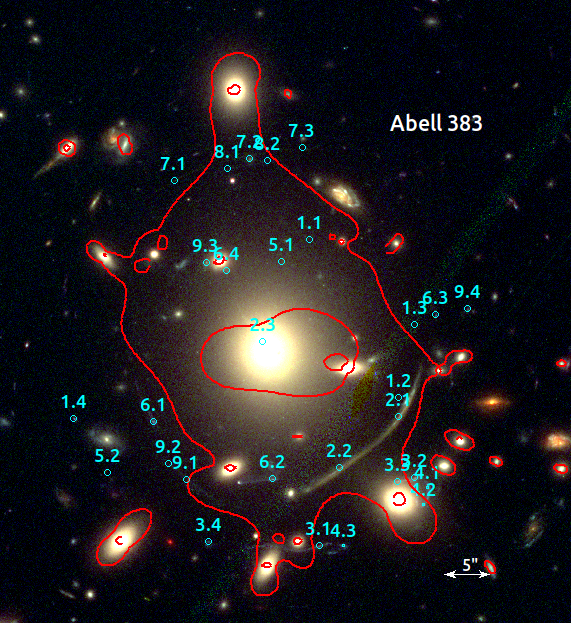}
        \includegraphics[width=0.5\linewidth,height=0.5\linewidth]{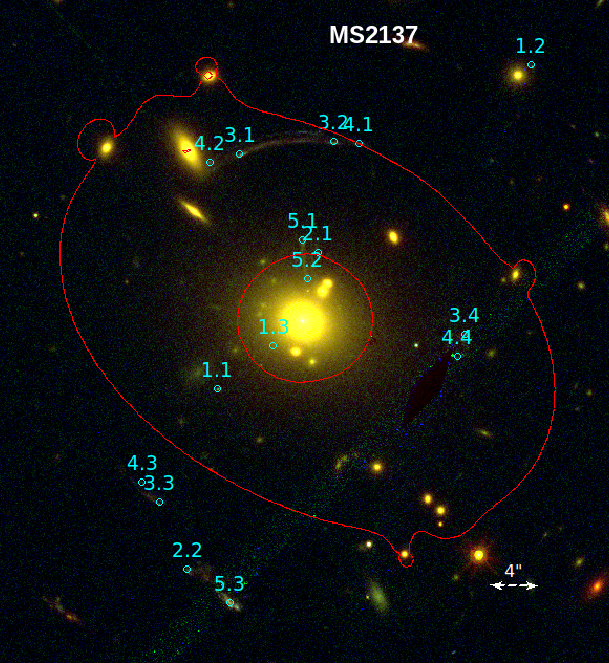}
        \includegraphics[width=0.5\linewidth,height=0.5\linewidth]{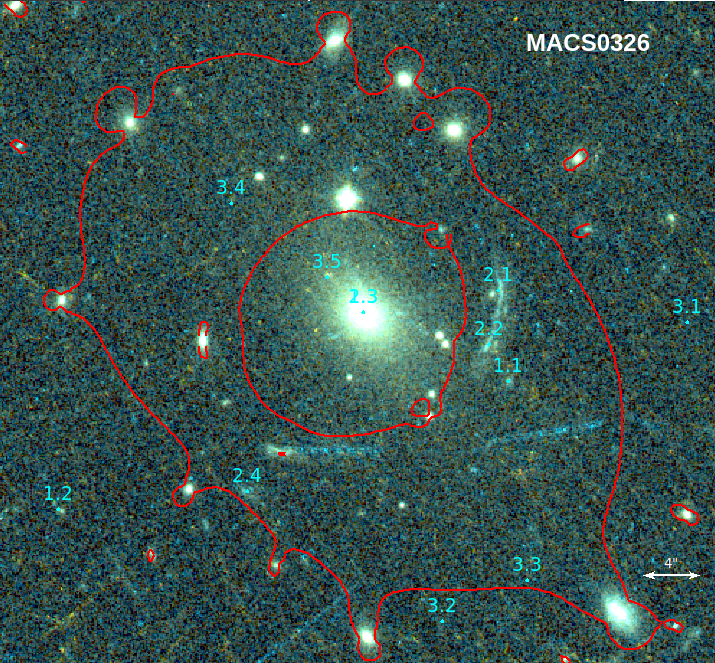}
        \includegraphics[width=0.5\linewidth,height=0.5\linewidth]{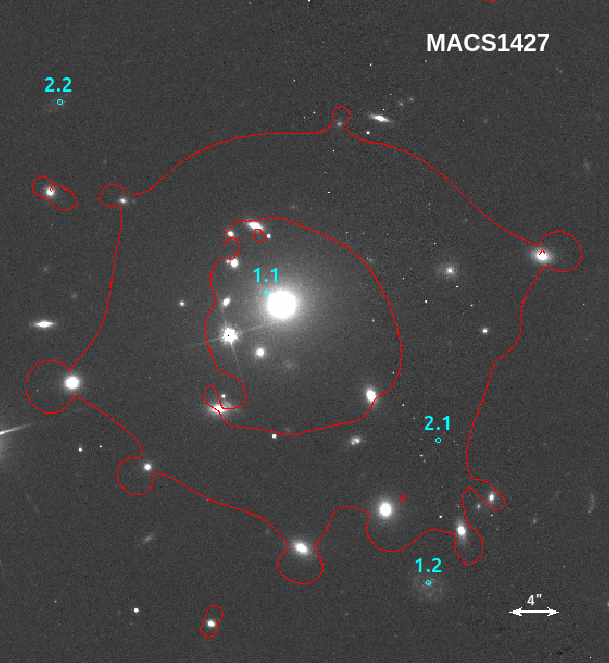}
        \caption{Images of the four clusters newly modeled in this paper. All images are oriented North-up, East-left. {\it Top Left:} \hst~~composite color image of \abell~created using a combination of WFC3/IR imaging in the red (F160W) and ACS imaging in the green (F814W) and blue (F606W). Multiply imaged galaxies are labeled in cyan. The red curve marks the location of the critical curve for a source at $z=3.0$. {\it Top Right:} False color image of \ms~created using a combination of WFC3/IR imaging in the red (F160W) and ACS imaging in the green (F814W) and blue (F606W). Multiply imaged galaxies are labeled in cyan. The red curve marks the location of the critical curve for a source at $z=2.5$.{\it Lower Left:} False color image of \mstwo~created using a combination of ACS imaging in the green (F814W) and blue (F606W). Multiply imaged galaxies are labeled in cyan. The red curve marks the location of the critical curve for a source at $z=3.0$. {\it Lower Right:} F814W image of \msthree. Multiply imaged galaxies are labeled in cyan. The red curve marks the location of the critical curve for a source at $z=1.2$. 
        }\label{fig.allmodels}
    \end{minipage}%
\end{figure*}

 One cluster-scale dark matter halo is used to model the cluster, and a total of four galaxy-scale halos are included to separately model the mass distribution of the BCG, as well as three cluster members located near systems 3 and 4. The best fit parameters for these halos are given in Table~\ref{tab.a383model}. The integrated density profile is presented in Figure~\ref{fig.densityprof}. A snapshot of the radial arc in the south-west portion of the cluster is shown in Figure~\ref{fig.giantarcs}. The cluster, multiple images, and the critical lines for the model at $z=3.0$ are shown in Figure~\ref{fig.allmodels}.


\subsubsection{\msfull}

 One cluster-scale dark matter halo is used to model the cluster, and two galaxy-scale halos are used to separately model the mass distribution of the BCG and one cluster member located near systems 3 and 4. The parameters for these halos are given in Table~\ref{tab.ms2137model}. 
The integrated density profile is presented in Figure~\ref{fig.densityprof}. 
A snapshot of the radial arc in the north-east portion of the cluster is presented in Figure~\ref{fig.giantarcs}. The cluster, multiple images, and the critical lines for the model at $z=3.0$ are shown in Figure~\ref{fig.allmodels}.


\subsubsection{\mstwofull}

One cluster-scale dark matter halo is used to model the cluster, one galaxy-scale halo is used to separately model the mass distribution of the BCG, and one additional halo is included for the bright cluster member located near system 3. The best fit parameters for these halos are given in Table~\ref{tab.ms0326model}. The integrated density profile is presented in Figure~\ref{fig.densityprof}. A snapshot of the two radial arcs near the BCG are shown in Figure~\ref{fig.giantarcs}. The cluster, multiple images, and the critical lines for the model at $z=3.0$ are shown in Figure~\ref{fig.allmodels}.


\subsubsection{\msthreefull}

 This cluster has not been mass modeled before. One cluster-scale dark matter halo is used to model the cluster and one galaxy-scale halo is used to separately model the mass distribution of the BCG. The best fit parameters for these halos are given in Table~\ref{tab.ms1427model}. The integrated density profile is presented in Figure~\ref{fig.densityprof}. A snapshot of the radial arc is shown in Figure~\ref{fig.giantarcs}. The cluster, multiple images, and the critical lines for the model at $z=3.0$ are shown in Figure~\ref{fig.allmodels}.


\subsubsection{MACS J0417.5-1154}

We refer the reader to \cite{jauzac2019} and \cite{mahler2019} for more  details regarding this mass model. The only modification we make to the model described in \citetalias{jauzac2019} is the introduction of the parameters listed in Table~\ref{tab.bcgphot} to constrain the mass halo for the BCG. A snapshot of the radial arc is presented in Figure~\ref{fig.giantarcs}. The integrated density profile is presented in Figure~\ref{fig.densityprof}.

\subsubsection{MACS J0949.8+1708}

\begin{figure*} 
    \centering
        \includegraphics[width=1.0\linewidth]{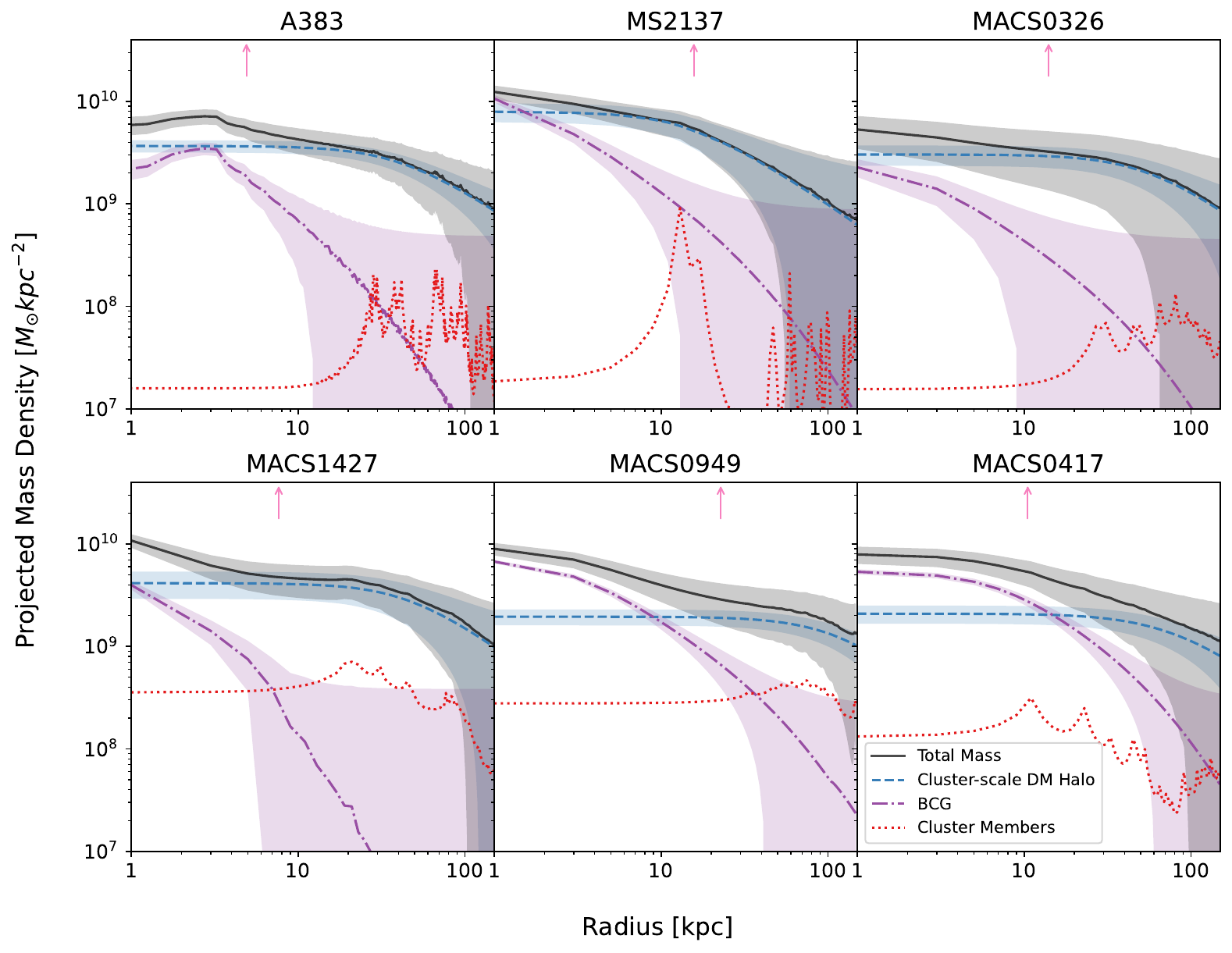}
        \caption{ The integrated density profiles for all six clusters studied in this work. The uncertainty in the measurement is plotted as the shaded region around the line, and only incorporates the uncertainty returned by \lenstool; systematic errors are not included. \color{black}
Uncertainty measurements are not included for the mass profile of the cluster member galaxies as they are too large to be meaningful. \color{black}~Density is measured in units of $\text{log}~(M_{\odot}$/$\text{kpc}^2)$. \color{black}The arrows denote the distance between the BCG and the barycentre of the nearest radial arc used as a constraint in the lens model.\color{black}}
        \label{fig.densityprof}
\end{figure*}

We refer the reader to \cite{allingham2023} for more  details regarding this mass model. The only modification we make to the model described in this paper is the introduction of the parameters listed in Table~\ref{tab.bcgphot} to constrain the mass halo for the BCG. A snapshot of the radial arc is presented in Figure~\ref{fig.giantarcs}. The integrated density profile is presented in Figure~\ref{fig.densityprof}.


\color{black}
\subsection{Lensing Model Prior Distributions} \label{sec.lenstoolpriors}
\begin{figure*} 
    \centering
        \includegraphics[width=0.4\linewidth]{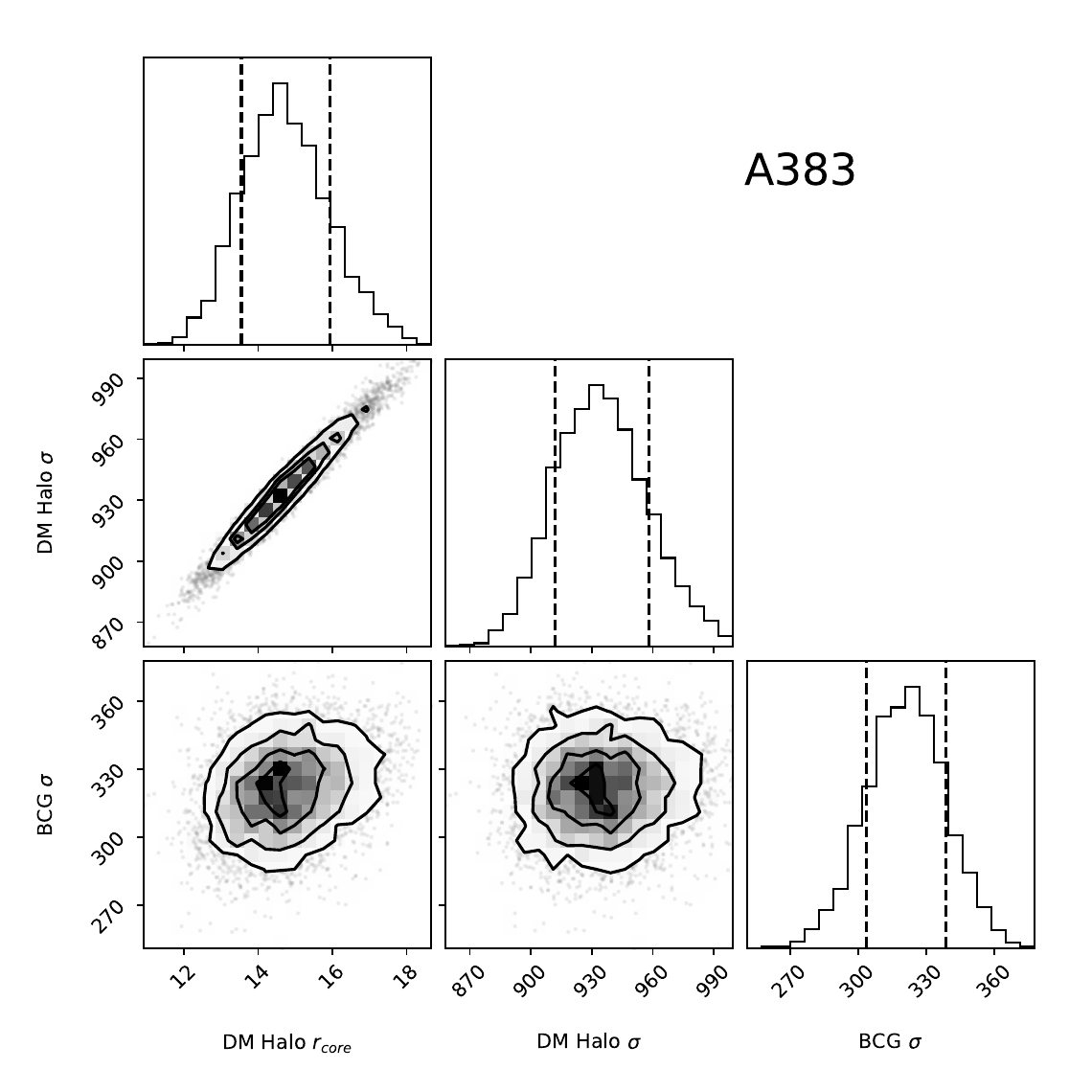}
        \includegraphics[width=0.4\linewidth]{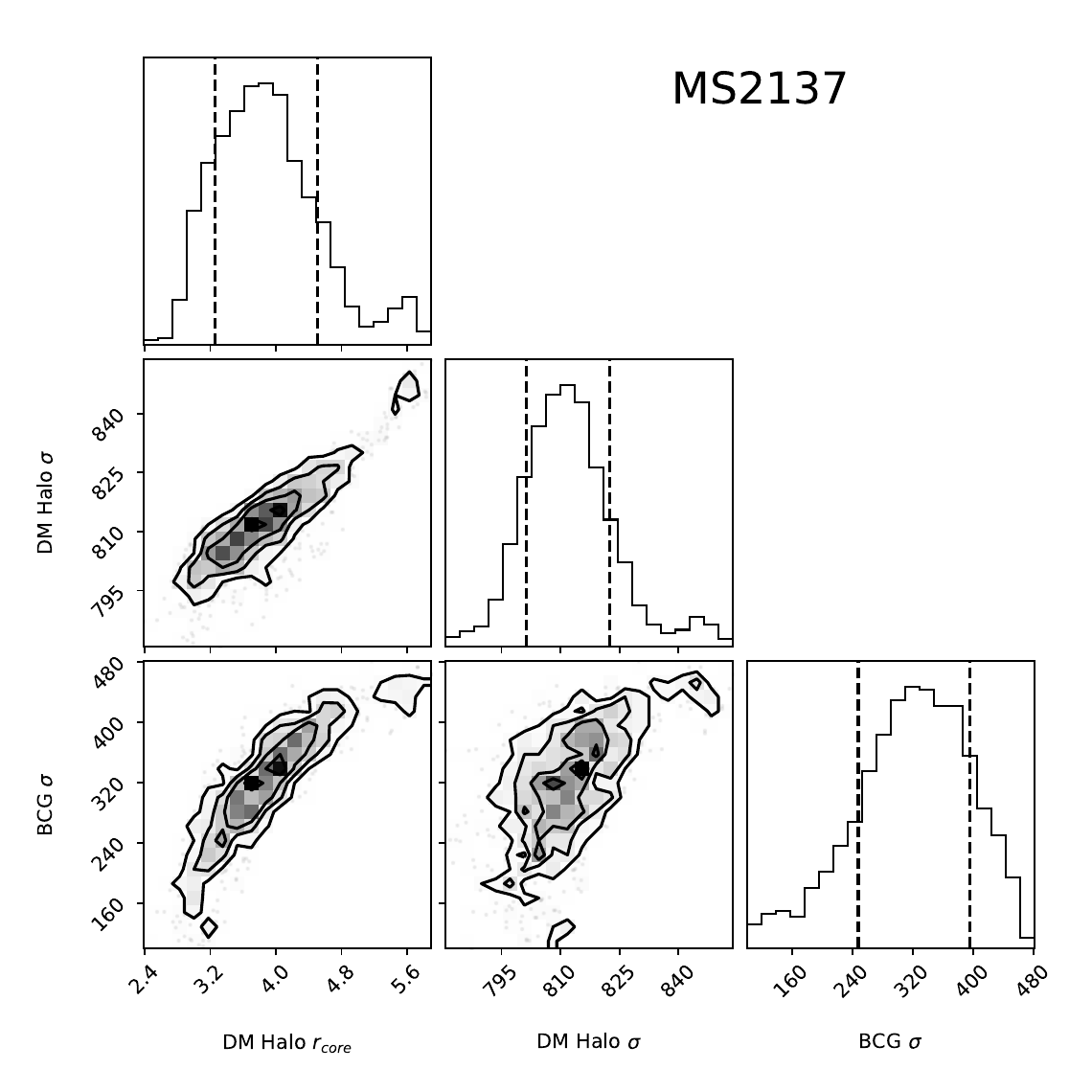}
        \includegraphics[width=0.4\linewidth]{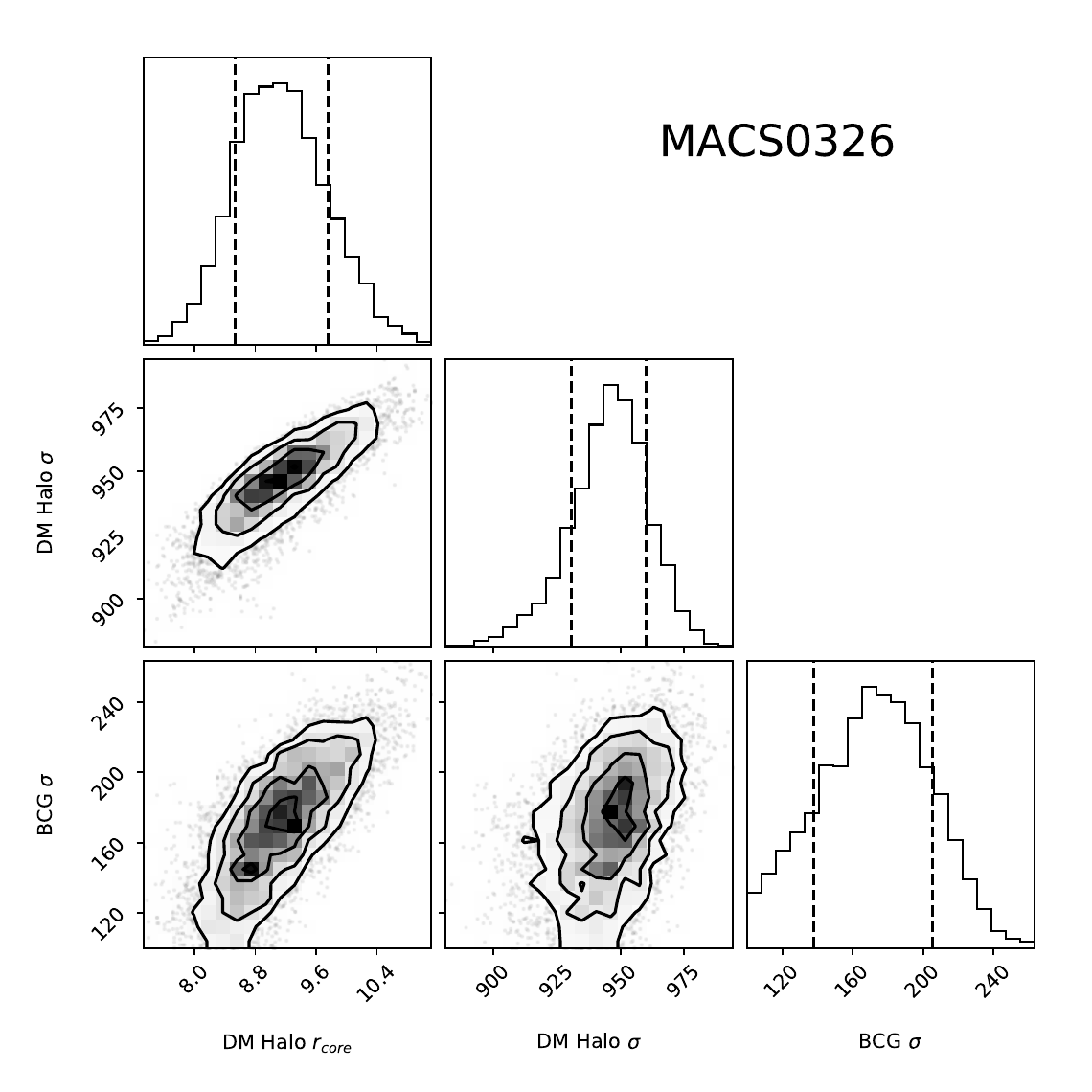}
        \includegraphics[width=0.4\linewidth]{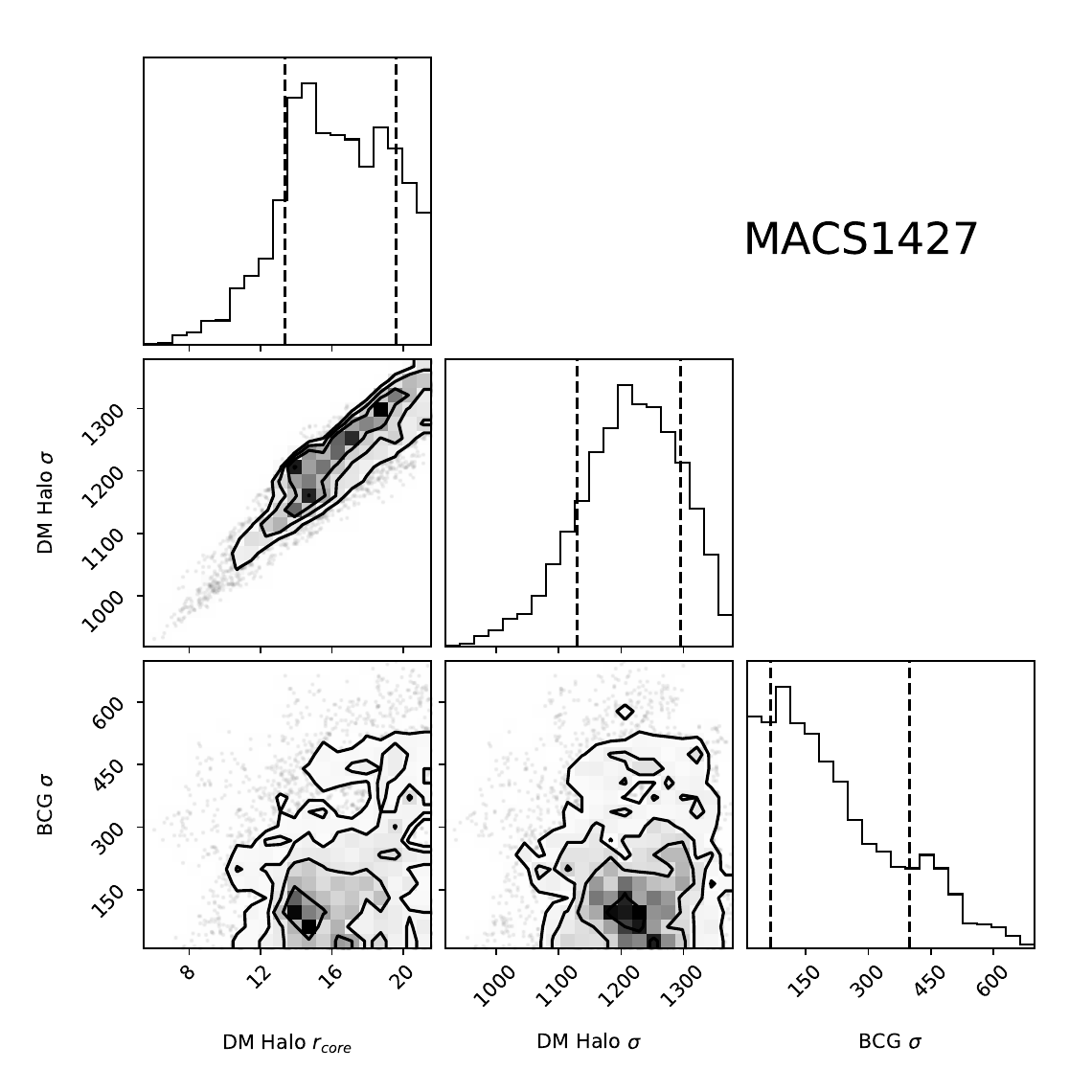}
        \includegraphics[width=0.4\linewidth]{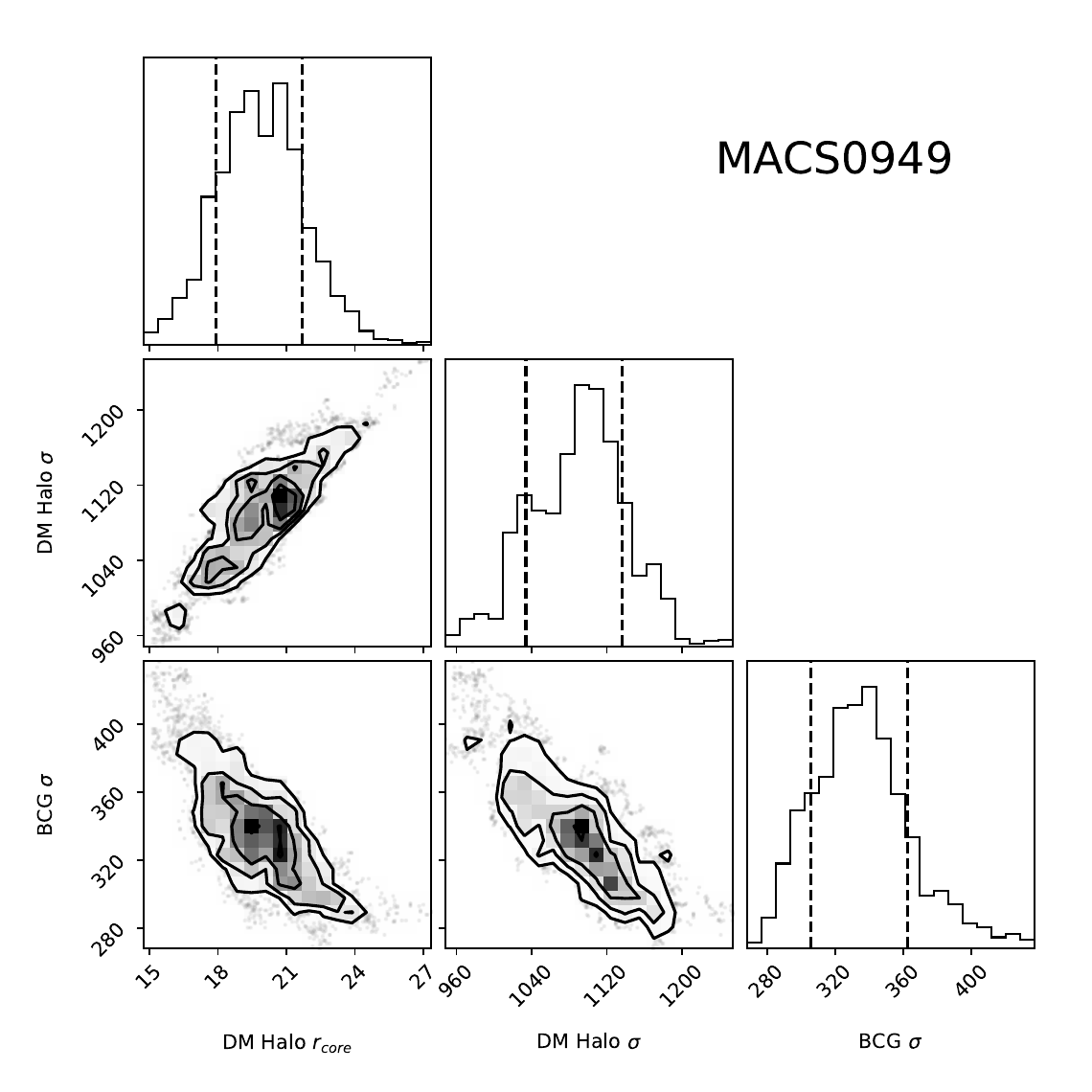}
        \includegraphics[width=0.4\linewidth]{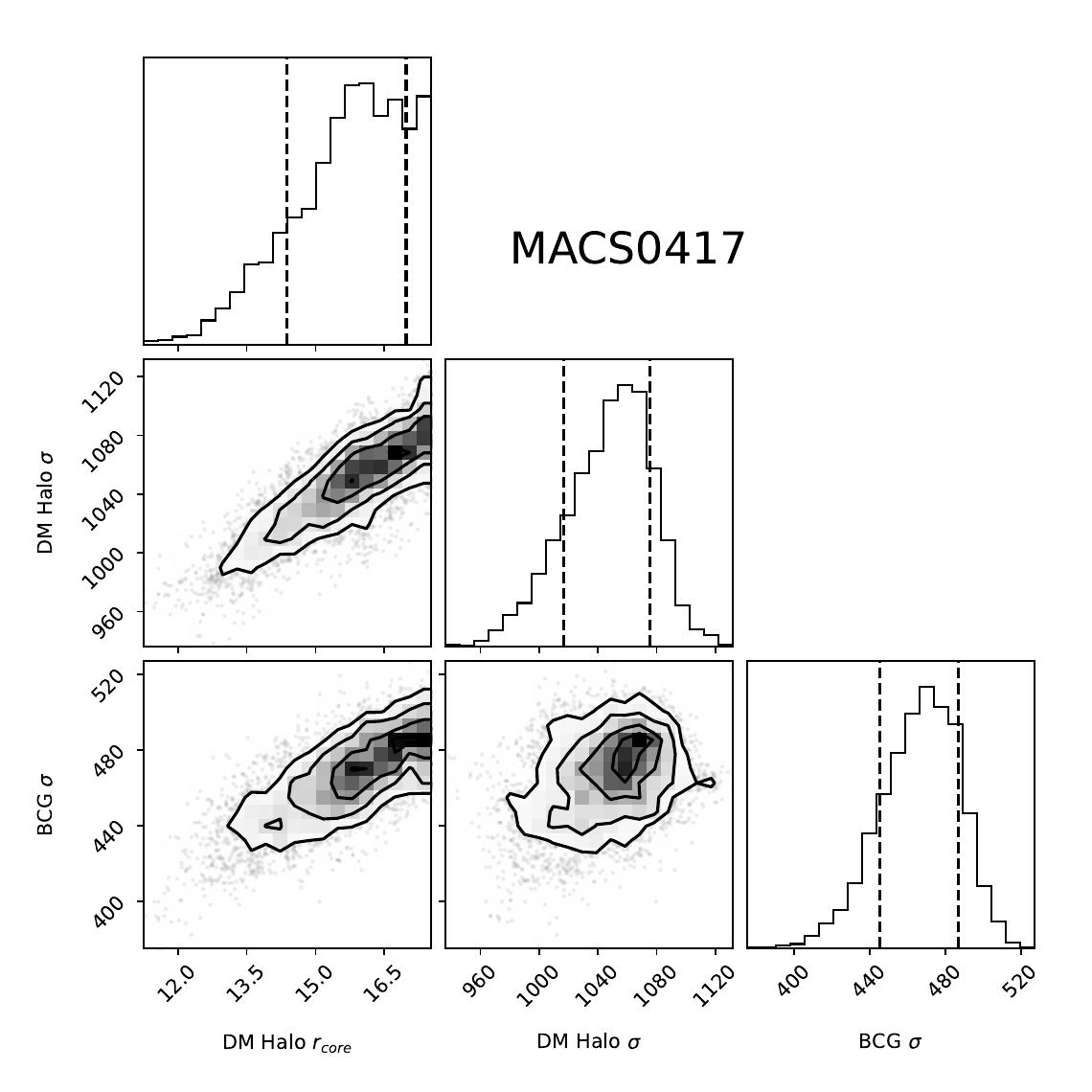}
        \caption{\color{black} Corner plots of the relevant parameters from the \lenstool~fitting procedure of the six clusters presented in this chapter. The dark matter halo of the cluster is denoted as the `DM' halo. The vertical lines mark the one-sigma region. }
        \label{fig.cornerplotch3}
\end{figure*}

The above models each contain different parameters that are optimized over a parameter space that differs slightly between clusters. The given parameter space for each parameter takes the form of a uniform distribution, which is the default distribution for priors in \lenstool~(\citealt{jullo2007}). This ensures that the full parameter space for each individual parameter is explored without biasing the search toward any particular value. The specific boundaries for each parameter space are provided in Table~\ref{tab.LTpriors}, and the corner plots for the key parameters in the fitting procedure are displayed in Figure~\ref{fig.cornerplotch3}. These parameters are the $r_{\mathrm{core}}$ parameter for the dark matter halo, and the velocity dispersions for both the dark matter halo and the BCG, which are essentially scaling factors for determining the mass distribution of the primary dPIE cluster-scale dark matter and BCG stellar mass halos within the lens models. There is a strong correlation between the velocity dispersion of the dark matter halo and the measurement for $\rcore$. Additionally, the lens model parameters are not very well constrained in \msthree, which is due to the lack of constraints within the lens model. The models for MACS J0949 and MACS J0417 are unchanged from their published versions, save for the addition of the explicitly constrained BCG, and are included here for completeness.


\begin{table}
\color{black}
	\centering
	\caption{Prior distributions for \lenstool~parameters optimized in the fitting procedure. The priors for $\Delta x$ and $\Delta y$ are given in units of arcseconds relative to the center of the cluster, which is fixed to the position of the BCG. The values in the prior column indicate the lower and upper bounds of the uniform prior assigned to each parameter. }
	\label{tab.LTpriors}
	\begin{tabular}{lccc} 
		\hline
		Parameter & Units & Prior \\
		\hline
  Cluster-scale dark matter dPIE halo & & \\
  \hspace{.2cm} $\Delta x$ & arcseconds & (-5, 5) \\
  \hspace{.2cm} $\Delta y$ & arcseconds & (-5, 5) \\
  \hspace{.2cm} $\epsilon $ & .. & (0, 0.8) \\
  \hspace{.2cm} $\theta$ & deg & (0, 180) \\
  \hspace{.2cm} $\rcore$ & kpc & (1, 100) \\
  \hspace{.2cm} $\sigma_0$ & km $\mathrm{s}^{-1}$ & (500, 1000)\\   
  BCG dPIE Halo & & \\
  \hspace{.2cm} \abell~$\sigma_0$ & km $\mathrm{s}^{-1}$ & (220, 420)\\ 
   \hspace{.2cm} \ms~$\sigma_0$ & km $\mathrm{s}^{-1}$ & (120, 500)\\ 
  \hspace{.2cm} \mstwo~$\sigma_0$ & km $\mathrm{s}^{-1}$ & (100, 350)\\ 
  \hspace{.2cm} \msthree~$\sigma_0$ & km $\mathrm{s}^{-1}$ & (100, 700)\\ 
  \hspace{.2cm} MACS 0949 $\sigma_0$ & km $\mathrm{s}^{-1}$ & (250, 450)\\ 
  \hspace{.2cm} MACS 0417 $\sigma_0$ & km $\mathrm{s}^{-1}$ & (350, 600)\\ 
  Cluster galaxy scaling & & \\
  \hspace{.2cm} $\sigma_*$ & km $\mathrm{s}^{-1}$ & (25, 250) \\ 
  \hspace{.2cm} $r_{\mathrm{cut}*}$ & kpc & (3, 50) \\ 
    dPIE halos of individually optimized galaxies \\
    \hspace{.2cm} $\epsilon$ & .. & (0, 0.8) \\
    \hspace{.2cm} $\theta$ & deg & (0, 180) \\ 
    \hspace{.2cm} $\rcore$ & kpc & (0, 1) \\
    \hspace{.2cm} $\rcut$ & kpc & (0,10) \\
    \hspace{.2cm} $\sigma_0$ & km $\mathrm{s}^{-1}$ & (10, 300) \\
    Unknown redshifts & & \\
    \hspace{.2cm} z & .. & (1,7) \\ 
    \hline

	\end{tabular}
\end{table}

\color{black}

\subsection{Combining Kinematic Profile with Lensing } \label{sec.kinlensing}

Stellar kinematics generally probe the total mass distribution, but in the case of galaxies, where significant amounts of stellar light are concentrated, the mass profile is strongly dominated by baryons. We can thus combine the kinematic measurements of the BCGs of these clusters, which effectively measure the baryonic mass, with strong lensing models, which are sensitive to dark matter, to disentangle the degeneracy between dark and baryonic matter. We elect to employ kinematic measurements of the BCG in our models to more accurately probe the mass distribution at the center of these galaxy clusters. We separately parameterize the BCG using a dPIE profile with \lenstool, using parameters derived from photometry. We leave the velocity dispersion as the only free parameter for this profile. We can then constrain the BCG velocity dispersion using our real kinematic measurements from MUSE. In this way, we can use physical measurements of the BCG to constrain the lens model. 

The velocity dispersion of the BCG is incorporated as a constraint into the model through a-posteriori analysis of the MCMC \lenstool~model. The original \lenstool~model is first modified to incorporate photometric information (discussed in Section \ref{sec.bcgphot}) for the BCG into the model. This is done by creating a separate mass halo to account for the BCG, with the parameters $r_{\mathrm{core}}$ and $r_{\mathrm{cut}}$ fixed to the values obtained from the surface brightness profile fit. Ellipticity and position angle are also fixed to the values from the {\texttt{GALFIT}} fitting. The remaining free parameter in the halo, $\sigma_{0}$, is left free as a proxy for the stellar mass to light ratio (\citealt{sand2004}, \citealt{sand2008}, \citealt{bergamini2019}). This parameter is given a prior that is informed by measurements of the stellar mass to light ratio.

Adding this information to the model allows us to directly calculate a model version of the velocity dispersion of the BCG based on Jeans fitting. \color{black}In this work, we assume that each BCG can be described by a spherical mass distribution, and we  do not explicitly incorporate velocity anisotropy into our models, as any effects from including this component do not have a large impact on the measured velocity dispersion (see the discussion in Section~\ref{sec:anisotropy}).\color{black}~The form of the spherical Jeans equation is thus as follows: 

\begin{equation*}
    \\
\sigma _{\mathrm{los}}^2(R) = \frac{2G}{\sum_{*}^{}} \int_{R}^{\infty } \frac{{}\nu_{*}(r)M(r)\sqrt{r^{2}-R^{2}}}{r^2}  dr,
\end{equation*}

\noindent where $\nu_{*}$ describes the three-dimensional profile and $\sum_{*}$ refers to the two-dimensional profile of the stellar component of the BCG. These profiles are drawn directly from the photometric fits to the BCG described in Section~\ref{sec.bcgphot}. $M(r)$ accounts for the total enclosed mass inside a radius, $r$, and must thus account for both the stellar and DM mass. We draw the value for the enclosed mass directly from the \lenstool~model at this stage, as the degeneracy between DM and stellar mass makes it difficult to avoid double-counting the mass of the BCG if we add it separately to the \lenstool~value. We can then calculate a model velocity dispersion for the BCG at any given radius, $R$. 

The optimization of the \lenstool~model with respect to the stellar kinematics is performed by adding the error from the model to the error of the stellar kinematics, 

\begin{equation*}
    \\
\chi^{2}_{\mathrm{VD}} = \sum_{i}^{} \frac{\left (\sigma _{i} - \sigma _{i}^{obs}  \right )^{2}}{\Delta ^{2}_{i}},
\end{equation*}

\noindent where $\Delta_{i}$ is the uncertainty in the observed velocity dispersion measurements\color{black}, and $i$ represents the radial bins where the velocity dispersion was measured. The optimized best-fit \lenstool~model is then the one that minimizes the difference of the model velocity dispersion within each radial bin relative to the measured velocity dispersion within those bins.\color{black}

The calculation of the velocity dispersion error is performed after the lensing minimization has been completed, separately from \lenstool. As a result, the velocity dispersion and lensing likelihoods can be treated as independent from each other, and the total likelihood is then the product of the velocity dispersion and lensing likelihoods, where we assume that both the lensing and dynamical models carry equal weight in the final calculation. The resulting 'best-fit' model is then simply the model that minimizes the total $\chi^2 = \chi^2_{SL} + \chi^2_{VD}$. This treatment of the likelihoods as independent quantities follows previous work done to combine dynamics and lensing models (e.g. \citealt{sand2004}, \citealt{newman2013density}).  

Maximizing the global likelihood function allows us to separate DM and baryons into distinct profiles, as seen in  Figure~\ref{fig.densityprof}. In doing so, we can compare the DM and baryon profiles (\color{black}blue\color{black}~and \color{black}purple/red\color{black}~lines, respectively) to the total mass distribution (black line) to see the relative contributions of each as a function of radius. Each of these profiles possess a core-like structure within the inner 50 kpc. This is perhaps not unexpected, given that all of these clusters have radial arcs, which are preferentially produced in mass distributions with a shallow inner slope. However, the agreement between the shape of the profile for each of these clusters is fairly significant, as it shows that the dark matter distribution in the center of the cluster is centrally concentrated.

\begin{figure} 
    \centering%
        \includegraphics[width=0.7\linewidth]{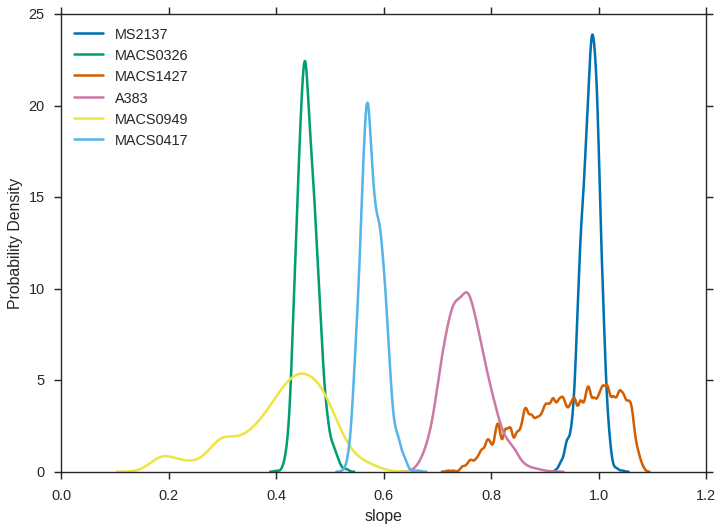}
        \includegraphics[width=0.7\linewidth]{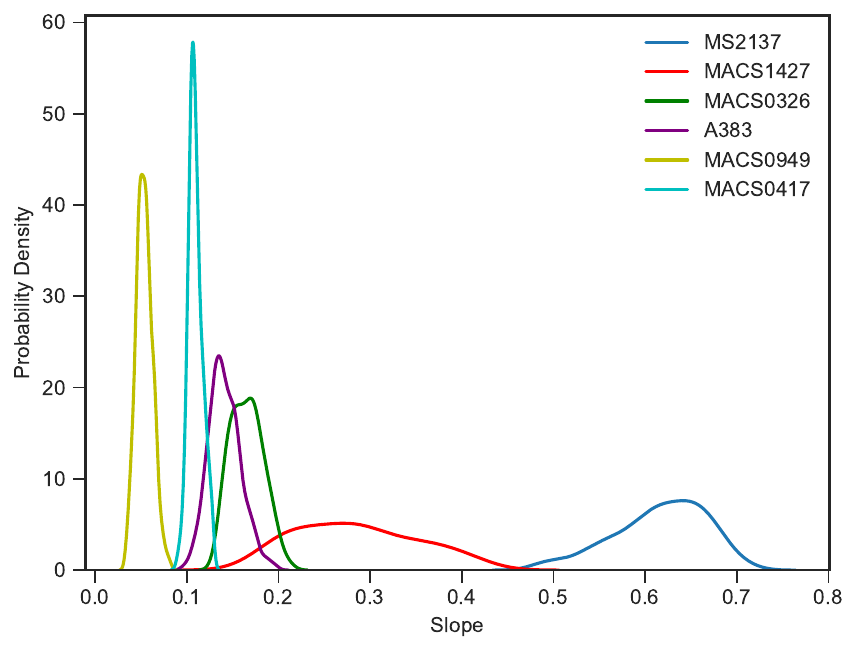}
        \caption{\color{black} \textit{Top}: \color{black} Density probability distribution for the slope measurements in all 6 clusters\color{black}, as calculated from the slope of the total mass density profile with the slope of the BCG subtracted. \color{black} The probability density is calculated from the MCMC chain run for each model and is based on the derived parameters for the cluster DM halo. \color{black} \textit{Bottom}: Density distribution for the slope of the parameterized dPIE cluster-scale dark matter halo in each cluster. \color{black}
        }\label{fig.pdf}
\end{figure}

\subsection{Cluster Dark Matter Density Profile}

We use a dPIE profile to construct \color{black}the\color{black}~ density profile \color{black}of the DM halo\color{black}~ for these clusters. Since we do not use the usual NFW profile, an examination of the properties of the dPIE profile is worth considering. 

The obvious test of this profile is to compare it against a `non-core' model.  In \cite{newman2013lens}, this comparison between a core and non-core model was performed by using a gNFW profile and a `cored' NFW profile, with the cored model being ultimately favored. 
In our case, we choose to use dPIE profiles over NFW because we lack weak lensing information for the clusters \msthree~and \mstwo. As shown in \cite{limousin2022}, weak lensing is crucial for placing reasonable priors on the scale radius of the NFW profile, and without these priors any attempts to use \lenstool~to create NFW profiles will result in models that are degenerate with their own parameters. As a result, until we are able to perform weak lensing analyses for these clusters and obtain priors on their parameters, the information we derive from NFW fitting will have a low statistical significance. While we can investigate weak lensing modeling, the \HST~data for these clusters are too shallow to perform a meaningful fit to the data (e.g. the number density of background sources is too low). Additionally, the lack of multiple bands makes weak lensing analysis difficult to incorporate as the contamination from foreground and cluster objects will be almost impossible to remove (\citealt{jauzac2012,niemiec2023}). Since \color{black}only two of the\color{black}~clusters covered in this work have published weak lensing models (see \citealt{newman2013density,jauzac2019,allingham2023}), we elect not to repeat this analysis since the available data is the same, and we are not introducing new weak lensing modeling techniques. Additionally, by not using an NFW profile to model the cluster DM profile, we avoid the degeneracy between the scale radius and the asymptotic gNFW profile slope that was demonstrated in \cite{he2020}, and do not bias our results toward low slope values as a result of measurement choices in a weak lensing profile. We leave the potential exploration of NFW profiles to future work.

Instead, we choose to create `core' and `non-core' dPIE models following the approach presented in \cite{limousin2022}. We perform this test by restricting the $r_{\mathrm{core}}$ radius to be smaller than 10 kpc to create `non-cored' models, and comparing the resulting models against our `cored' dPIE models, where the core radius was left as a free parameter. The results of this experiment are definitive. In each case, the $\chi^2$ value of the `non-cored' model increased significantly in comparison to the cored model. For instance, the `cored' model for MACS J0326 has a $\chi^2$ of 23.4, with an rms of 0.37". In comparison, the `non-cored' model has a $\chi^2$ of 179, with an rms of 3.3", which indicates that the model is unable to reproduce the observational constraints. This discrepancy is consistent across all six clusters, indicating that 'non-cored' models are not favored as a solution.


\section{Discussion} \label{sec:disc}

Strong lensing is a powerful tool for examining the inner slope of the dark matter density profile, which can then be used to compare with, and improve upon the information used to build CDM simulations. The clusters modeled in this paper all have radial arcs, which are uniquely suited for providing more precise constraints in the inner part of the cluster, i.e. near the BCG.

\subsection{How Rare are Radial Arcs?}
Radial arcs are still relatively uncommon among lenses, both because the geometric requirements for their appearance are so stringent, and then even when they do appear, they can easily be obscured by the light of surrounding galaxies and the BCG \citep{matthias2010}. Visual examination of Figure~\ref{fig.giantarcs}, for example, would not immediately make the presence of a radial arc obvious; it is only through spectroscopic confirmation that we can definitively say an arc is present and assign it a redshift, which defines its power as a constraint. As a result, building a large sample of clusters with radial arcs is not trivial. The six clusters selected in this paper are chosen from a dataset of around 150 MUSE cubes (PID 0104.A-0801; PI A. Edge) from the Kaleidoscope survey, a large `filler' program, and are selected from within that dataset specifically for the presence of radial arcs. The ratio of clusters with radial arcs vs clusters without these arcs in the Kaleidoscope sample is around \color{black}4\color{black}\%, which easily shows that radial arcs, while a powerful constraint on the inner DM profile, are a very unique physical feature and cannot be relied upon to appear in lensing analyses of cluster mass profiles at scale. 

An additional, equally important consideration for the robustness of our measurements is that the model and slope profiles presented in this work exclusively represent clusters that contain radial arcs. This naturally means that all of these clusters are more core-like, as total density distributions steeper than $\rho \propto r^{-2}$ do not produce radial arcs (see {\citealt{hattori1999}). As a result, we incur another degree of selection bias in our results by excluding systems that might be more cuspy. While we do plan to examine this in the future using other clusters in the Kaleidoscope survey, it is useful to mention what potential impact this bias might have by referring to previous work done by \cite{sand2004} and \cite{newman2013lens} on A383 and MS2137.

In each of these papers, these two clusters were a part of a larger sample, which included clusters that had tangential arcs but did not have radial arcs. The authors could thus perform a comparison of their results relative to clusters without radial arcs, and the shallowness of the inner DM density profiles for both A383 and MS2137 did not differ significantly from the profiles of clusters with only tangential arcs. These results are encouraging, and we plan to more thoroughly test the robustness of our own sample in future work. In this paper, we present our models as a specific study of density distributions for clusters with radial arcs and shallow IFU spectroscopy, rather than an examination of the general shape of the DM density profile for all clusters. Despite these limitations, however, we are able to obtain measurements for the inner DM slope that are both consistent with previous mass models and are in agreement with other observational measurements from literature, thus demonstrating the strength of using radial arcs as lensing constraints in tandem with MUSE spectroscopy. 


\subsection{Comparison with previous work}


\subsubsection{\abellfull}\label{a383densprofref}

\abell~has been modeled numerous times, due to its properties and fortuitous magnification of a background $z\sim6$ galaxy. The most recent model was created using a combination of strong and weak lensing, and was part of a compilation of models built using the complete sample of CLASH clusters (\citealt{zitrin2015}; Z15). The multiple image systems used in \citetalias{zitrin2015} correlate to those used in this paper, with the exception of our system 1, which is a new detection. System 1 in \citetalias{zitrin2015} is associated here with our system 2; the remaining systems are numbered in accordance to \citetalias{zitrin2015}. We note that our system 2 is a combination of \citetalias{zitrin2015} system 1 and system 2, as we treat the tangential arc and the radial arc as the same system based on our MUSE observations. 

Systems 1-5 are fixed to spectroscopic redshifts, where systems 1, 2, and 5 are measured from our MUSE spectroscopy, and are identical to the values presented in previous literature, including \citetalias{zitrin2015}. Systems 3 and 4 are not detected in our MUSE data. We thus fix these to the redshift $z=2.55$, a spectroscopic measurement obtained by \cite{newman2011}. The remaining redshifts are derived by our model, and they generally tend to be lower, but within $\pm~0.5$ of the values found by \citetalias{zitrin2015}, with one exception. System 8 is a more severe underestimation at~\color{black} a model redshift of\color{black}~$z=1.746$ compared to~\color{black} a model redshift of\color{black}~$z=3.1$ from \citetalias{zitrin2015}.

Despite the difference in \color{black}best-fit redshifts\color{black}, our mass estimates are generally in agreement. In \citetalias{zitrin2015}, the mass enclosed in 100 kpc is $\sim6 \times 10^{13} M_{\odot}$. In our model, we find that the mass enclosed in the same radius is $\sim5.5 \pm 0.06 \times 10^{13} M_{\odot}$. The other major model for this cluster that we reference in this paper is from  \cite{newman2013lens}, which finds an enclosed mass within 100 kpc of $\sim6 \times 10^{13} M_{\odot}$. Our mass estimate thus matches well with the most recent parametric and light-traces-mass lens models of this cluster.

The error in the mass estimate is larger in the outskirts of the cluster as opposed to the inner regions, consistent with expectations from strong lensing models that are expected to be most accurate in the region enclosed by the critical curve. The total $\chi^2$ error estimate is $\sim54$, as well, which correlates to the mass profile error, and an rms of 0.53".   

The integrated density profile shown in Figure~\ref{fig.densityprof} is measured by summing the value of all pixels encapsulated within an annulus of width $r(n)-r(n-1)$, where $n$ corresponds to the step number.

\subsubsection{\msfull}

\ms~has similarly been modeled several times since the work by \cite{sand2008}. The most recent model is again from \cite{zitrin2015}. We break system 5, the large tangential arc, into two different pieces in order to improve the resolution of the model. Our systems 3 and 4 correspond to system 1 in \citetalias{zitrin2015}. System 2 in \citetalias{zitrin2015} corresponds to our system 5, and system 3 in \citetalias{zitrin2015} corresponds to our system 1. We add one new detection to the model: system 2, which we confirm via MUSE spectroscopy to be at $z=1.191$. This system corresponds to the second radial arc near the BCG, which had been noted before in photometry but lacked a spectroscopic detection both of itself and of a counter image that would make it viable to include in the lens model.

The most direct comparison we can make between our model and that of \citetalias{zitrin2015} is the mass estimate. \citetalias{zitrin2015} reports a total 2D integrated mass of $M(R<100~\mathrm{kpc})\sim4\times10^{13}M_{\odot}$.
A similar estimate of our mass within the same radius yields a value of $3.6 \pm 0.1 \times 10^{13} M_{\odot}$

The error in the mass profile is relatively small since our model is well-constrained by the inclusion of five systems with spectroscopic redshifts, with a total $\chi^2$ error estimate of $\sim40$ and an rms of 0.67".

\subsubsection{\mstwofull}
This is the first published strong lensing model for \mstwo, as well as the first mass estimate. Each system used to build the model has a spectroscopic redshift, which contributes to the reduction of overall systematic errors \citep{johnson2014}. Four multiple images out of the twelve used in the model are predictions made by the model (see Table~\ref{tab.ms2137arcs}). The redshift used for each system is fixed to the spectroscopic redshift of the arc with the highest S/N ratio, and is constrained to four significant figures. We report a total integrated mass of $M(R<100~\mathrm{kpc})\sim6\times10^{13}M_{\odot}$.
The $\chi^2$ value found for the model is $28.44$, while the overall rms in the image plane is $0.77"$. The largest contributors to the rms are images 1.1 and 2,4, which are predicted images. Because the model is constructed using only two shallow \hst~bands (each band has a short exposure time of 500 s; see Table~\ref{tab.hst}), it is somewhat difficult to do the usual color and morphology comparison typically used to identify other strongly-lensed galaxies. As a result, the positions of the predicted images are merely predictions, and are subject to change if deeper spectroscopic or photometric observations are acquired for this cluster.

\subsubsection{\msthreefull}
This is the first published strong lensing model for \msthree, and is also the first examination of this cluster in the visual band. However, it was discussed in \cite{ebeling2010} as a part of a survey of X-ray bright clusters from the MACS survey using {\it Chandra} data. The \hst~data used for the cluster is archival, and was observed as part of a SNAP project (PID 12884; PI: Harald Ebeling). Although only one \hst~band makes it difficult to identify multiple images based on the usual criteria of color and morphology, using it as a spatial reference alongside the MUSE redshifts makes it possible to create a basic lens model for the cluster. We report a total integrated mass of $M(R<100~\mathrm{kpc})\sim8\times10^{13}M_{\odot}$. We obtain a total $\chi^2$ of $\sim43$ with an rms of 0.81". Better quality imaging will almost certainly change the details of the model, but the current iteration still provides valuable information about the general shape of the mass in the cluster. Because the goal of this paper is to present lens models constructed mostly based on MUSE data, it is outside the scope of the current work to perform a thorough comparison of the mass profile against the existing {\it Chandra} data used in \cite{ebeling2010} for this cluster. However, deeper \hst~imaging and a subsequently improved lens model would make such a comparison more robust, and is thus left for future work.

\subsubsection{MACS J0417}

The lens model used for this cluster is identical to that presented in \citetalias{jauzac2019}. There are two BCGs in this cluster, but only the southern one has a radial arc. In \citetalias{jauzac2019}, both BCGs are separately parameterized with individual dPIE halos. In this work, we fix the parameters of the southern BCG to the properties we derive from our photometric measurements (see Table~\ref{tab.bcgphot}). The second change we make from the \citetalias{jauzac2019} model is to impose an additional selection criterion on the best-fit model from MCMC chain through our kinematic constraints. The resulting model does not differ strongly from \citetalias{jauzac2019} in the overall mass distribution. The reduced $\chi^2$ for the model in this paper is $\sim0.91$, compared to $\sim0.9$ from \citetalias{jauzac2019}, and the rms is 0.41", compared to 0.38" from \citetalias{jauzac2019}. Furthermore, the enclosed mass within 200 kpc is measured to be $1.8 \pm 0.04 \times 10^{14} M_{\odot}$, which is in agreement with the measurement from \citetalias{jauzac2019} from $1.78 \times 10^{14} M_{\odot}$. This also indicates that the model is in agreement with \cite{mahler2019}, which has similar values for the mass, rms, and $\chi^2$. The main change is in the value of the inner density slope, which we discuss more in Section~\ref{sec:innerprof}. In the lensing only model, this value is $\gamma ~\sim 0.55$, whereas in the lensing and BCG kinematics model the slope is $\gamma ~\sim 0.6$. These values still fall within the FWHM of the PDF, however.

\subsubsection{MACS J0949}

The lens model for this cluster is identical to that presented in \cite{allingham2023}. The authors use a separate parameterization of the BCG as well, so the only change we make to the model is to fix these parameters to the properties we derive from our photometric measurements (see Table~\ref{tab.bcgphot}). The enclosed mass within 200 kpc found by \citeauthor{allingham2023} is $2.0 \times 10^{14} M_{\odot}$, while the model in this paper finds a value of $2.1 \pm 0.07 \times 10^{14} M_{\odot}$ in the same radius. \citeauthor{allingham2023} finds a $\chi^2$ of 4.71 with an rms of 0.15", while the kinematic model we use finds a $\chi^2$ of 8.4 with an rms of 0.3". The model could likely be improved with a finer resolution of bins inside the MUSE cube for our measured velocity dispersions.

\subsection{Inner Dark Matter Density Profile}\label{sec:innerprof}

While strong lensing models measure the total density near the critical curves, lensing alone cannot independently distinguish between the contributions of baryonic and dark matter. Breaking this degeneracy is crucial to understanding the physics occurring at the center of these clusters. To that end, we introduce stellar kinematics into the model to constrain the effect of the baryonic mass of the BCG on the overall density distribution. We restrict our analysis to the kinematics of the BCG exclusively, as this is by far the most dominant stellar component in this region. Following the procedure laid out in \cite{newman2013density}, we measure the inner slope $\gamma = -d~\mathrm{log}~\rho_{\mathrm{DM}} ~/~d~\mathrm{log}~r$ over the range $r/r_{200} = 0.003-0.03$, which roughly corresponds to a range of 5 - 50 kpc.

\begin{figure} 
    \centering%
        \includegraphics[width=1.0\linewidth]{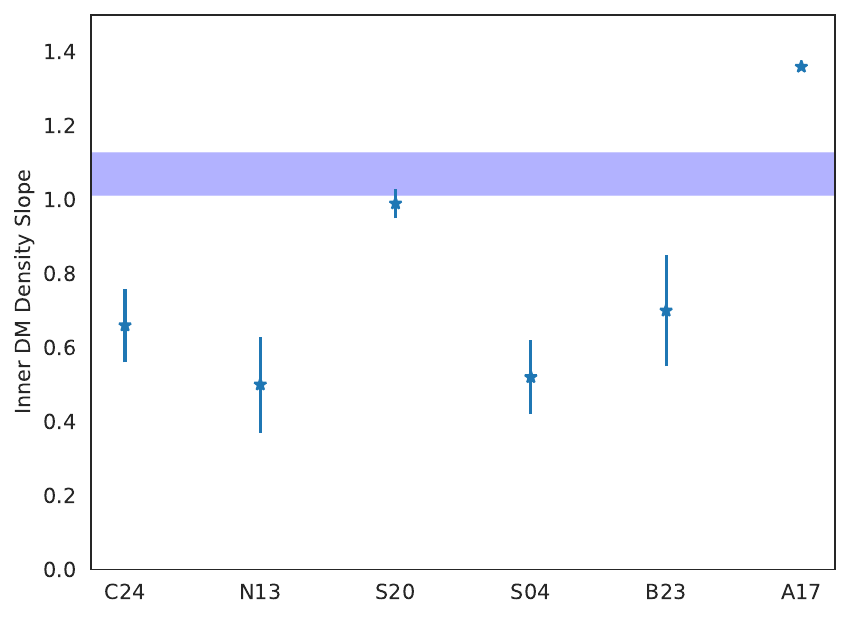}
        \caption{The average inner dark matter density slope as measured from five different papers, plotted from left to right as: this work (an average of six clusters), \protect\citetalias{newman2013lens} (an average of seven clusters), \protect\cite{sartoris2020} (Abell 1063), \protect\cite{sand2004} (an average of four clusters), \protect\cite{biviano2023} (MACS J$1206.2-0847$),  and \protect\cite{annun2017} (MACS J$0416-2403$). The expected slope measurement from the C-EAGLE simulations as reported in \protect\cite{he2020} is plotted as the blue shaded region.}
        \label{fig.papercompare}
\end{figure}

 We select these boundaries because the profile is strongly dominated by baryonic mass within 5 kpc, so the profile will always increase sharply within this region and is thus not a good indication of the shape of the DM density profile (see \citealt{newman2013density}), and outside of 50 kpc, the profile is dominated by dark matter (e.g. Figure~\ref{fig.densityprof}). The intermingling between baryonic and dark matter is thus best probed in the interior of this region. \color{black}The resulting measurements of the slope of the lens models within this range are shown in two ways in Figure~\ref{fig.pdf}. The primary results used in this work to estimate the slope measurements are plotted in the top portion of this figure, which shows the inner slope as measured from the total mass density of the lens model with the stellar component of the BCG subtracted. We choose to use this method to measure the inner slope because directly measuring the slope from the parameterized cluster-scale dPIE halo leads to results that may be misleading. The bottom component of Figure~\ref{fig.pdf} shows the measurement of the inner slope across the same range using the DM halo. The resulting slope values are almost universally cored. This correlates with the results presented in Figure~\ref{fig.densityprof}, which show near-universally flat dPIE halos in all 6 clusters. However, the shape of the dPIE halo does not begin to match the total mass density profile until after roughly 20-30 kpc, which is the cutoff threshold for the region where the BCG stellar mass should dominate. The measurements of the cluster-scale DM halo within the lens model are thus not necessarily a good reflection of the actual shape of the dark matter in the center of the cluster, especially because additional mass components that are not explicitly constrained within the lens model may be affecting the shape of the mass profile, such as intra-cluster light. As a result, rather than assuming that the cluster-scale DM halo represents the `true' distribution of dark matter in the cluster, we use the total mass obtained from the strong lensing models, which is a good measurement of the mass of the cluster regardless of the state of the individual components that make up that mass distribution, and subtract the mass of the BCG, which we explicitly parametrize using the photometric and kinematic measurements described in this work, to obtain the inner slope measurements we report for each of these clusters. \color{black}~
 The overall results suggest that each of these profiles is very cored, with an average slope measurement of $\gamma\sim0.66$ for all six clusters. Of all six clusters, \msthree~is the most poorly constrained, and could benefit the most from additional observations. However, its probability density is still centered firmly below 1.0, which suggests that it is more likely to be a cored cluster than a cuspy cluster. 

These results, when taken in aggregate, demonstrate that clusters with radial arcs \color{black}tend to\color{black}~present core-like density profiles. This is in-line with our current physical understanding of lensing, which requires the geometric shape of the mass distribution needed to produce radial arcs to be more \color{black}core-like.\color{black}

\section{Conclusions and Summary} \label{sec:conc}
We select four clusters with radial arcs from a total sample of 150 clusters observed with MUSE, and construct strong lensing models for these clusters, using at least two systems with spectroscopic redshifts from MUSE in each model. The radial arcs impose an additional constraint on the mass distribution near the center of the cluster, enabling a more precise examination of the inner density profiles for these clusters than is typical of strong lensing, as the relative scarcity of radial arcs means that these profiles are usually extrapolated into the inner region of the cluster. We add another two clusters with radial arcs that have already been modeled to our sample, for a total of six clusters with radial arcs. We note that radial arcs are preferentially produced in mass distributions with a shallow inner slope, and each of these clusters has a demonstrably shallow slope measurement regardless of whether photometric and kinematic measurements of the BCG are incorporated into the lensing model; however, including a central kinematic probe reduces systematics of the model. 

We additionally utilize stellar kinematics from the BCG to impose further constraints on the inner density profile following the methods used in \citealt{newman2013density}. This allows us to constrain the contribution of the baryonic mass to the overall cluster density profile. We obtain a mean dark matter slope value of $\gamma\sim0.66$ for all clusters, which is higher but generally consistent with the measurements from N13 ($\gamma\sim0.5$). Figure~\ref{fig.papercompare} places our results into context with several other observational papers, as well as with the results from the C-EAGLE simulations in \cite{he2020}. Of the five additional observational papers presented in the figure, \citetalias{newman2013lens} and \cite{sand2004} follow the methods presented in this paper to obtain their slope measurements. \cite{annun2017} uses a blend of X-ray, strong lensing, and BCG kinematics to obtain their measurement, although they assume a single power-law mass profile, which may over-estimate the inner slope if the 3D mass profile steepens with radius. \cite{sartoris2020} and \cite{biviano2023} exclusively use cluster member kinematics to obtain their slope measurements. These disparate values and techniques clearly show the observational tension currently facing this measurement. 

These results were typically obtained with shallow exposure times and in below-average observational conditions, which demonstrates that these techniques can be used to gain a general understanding of the shape of cluster density profiles without the need for deeper, more observationally constrained \HST~or spectroscopic data. However, we emphasize that our models and results can only improve with the addition of these types of data, and while the overall shape of our profiles should not change drastically, the overall measurements we obtain may shift. This can be seen in the difference of the work performed in \citeauthor{sand2008} and \citetalias{newman2013lens}. In \citeauthor{sand2008}, the authors derive a mean slope value of 0.5 for A383 and MS2137 using archival \hst~imaging and spectroscopy from Keck, whereas in N13, which used deeper \hst~images at a better wavelength for kinematic analysis, and deeper spectroscopy from Keck, the mean slope value for these two clusters is closer to 0.85. This shift is an example of the sort of change that we may expect to see if our models are ever redone in the future with, in particular, deeper observational data, which will allow for more reliable identification of multiply-imaged systems. 

Future work on this subject will make use of the techniques employed in this paper, applied on other galaxy clusters from the Kaleidoscope survey. Further analysis will include clusters without radial arcs and will exploit the Kaleidoscope sample to fully benefit from the uniform depth and selection of clusters that characterizes the survey. 

\section*{Acknowledgements}
CC, AN, and DJL acknowledge support from the UKRI FLF grant numbers MR/S017216/1 + NR/X006069/1. CC is additionally supported by a Durham Doctoral Studentship (Faculty of Science), awarded in 2020. MJ is supported by the United Kingdom Research and Innovation (UKRI) Future Leaders Fellowship `Using Cosmic Beasts to uncover the Nature of Dark Matter' (grant numbers MR/S017216/1 + NR/X006069/1). DJL is also supported by the Science and Technology Facilities Council (STFC) grants ST/T000244/1 and ST/W002612/1. This research made use of Photutils, an Astropy package for detection and photometry of astronomical sources (\citealt{bradley2022}).

\section*{Data Availability}

All \hst~data used for this work are available on the MAST
archive. All VLT/MUSE data used for this work are available on the ESO Science Archive. The lensing mass models and other data products will be shared by the authors upon request.


\bibliographystyle{mnras}
\bibliography{b.bib} 



\newpage
\appendix
\section{Lens Model Properties and MUSE Catalogue}

\begin{table*}
	\centering
	\caption{Parameters for the best-fit lens model of \abell. Error bars correspond to $1\sigma$ confidence level as inferred from the MCMC optimization. The parameters are provided for the cluster dark matter halo (dPIE DMH), the BCG halo (dPIE BCG), and each of the galaxy-scale perturbers included in the model (Perturber 1-3). $\Delta$R.A. and $\Delta$Decl. are defined in relation to the BCG, located at R.A.=2:48:03.37 and $\mathrm{Decl}.$=-3:31:45.29. Position angles are measured north of west, and the ellipticity $\epsilon$ is defined as $(a^2-b^2)/(a^2+b^2)$. $r_{cut}$ is fixed to 1000kpc for the cluster dark matter halo.}
	\label{tab.a383model}
	\begin{tabular}{lccccccc} 
		\hline
		Object & $\Delta$R.A. (") & $\Delta$Decl. (") & $\epsilon$ & $\theta$ ($^\circ$) & $r_{\mathrm{core}}$ (") & $r_{\mathrm{cut}}$  (") & $\sigma$ (km $s^{-1}$) \\
		\hline
dPIE DMH    & 0.43  & 2.26   & 0.17 & 106.66 & 50.80 & 1000.  & 878.22 \\
dPIE BCG   & 0     & 0      & - & - & -  & - & 325.59 \\
Perturber 1    & 14.69 & -16.10 & 0.79 & 96.31 & 0.31  & 5.89   & 189.58  \\
Perturber 2     & -0.17 & -23.49 & 0.37 & 139.62   & 0.89 & 5.00   & 139.13 \\
Perturber 3     & 3.27  & -20.82 & 0.77 & 92.84  & 0.81  & 2.00   & 290.66 \\
		\hline
	\end{tabular}
\end{table*}

\begin{table*}
	\centering
	\caption{Parameters for the best-fit lens model of \ms. Error bars correspond to $1\sigma$ confidence level as inferred from the MCMC optimization. The parameters are provided for the cluster dark matter halo (dPIE DMH) and the BCG halo (dPIE BCG). $\Delta$R.A. and $\Delta$Decl. are defined in relation to the center of the BCG, located at R.A.=21:40:15.16 and Decl.=-23:39:40.09. Position angles are measured north of west, and the ellipticity $\epsilon$ is defined as $(a^2-b^2)/(a^2+b^2)$. $r_{cut}$ is fixed to 1000kpc for the cluster dark matter halo.}
	\label{tab.ms2137model}
	\begin{tabular}{lccccccc} 
		\hline
		Object & $\Delta$R.A. (") & $\Delta$Decl. (") & $\epsilon$ & $\theta$ ($^\circ$) & $r_{\mathrm{core}}$ (") & $r_{\mathrm{cut}}$  (") & $\sigma$ (km $s^{-1}$) \\
		\hline
dPIE DMH    & 0.24  & 0.012 & 0.35 & 148.77 & 13.51 & 1000.& 784.81 \\
dPIE BCG    & -9.53 & 14.13 &- & -  & -  & - & 250.3 \\
		\hline
	\end{tabular}
\end{table*}

\begin{table*}
	\centering
	\caption{Parameters for the best-fit lens model of \mstwo. Error bars correspond to $1\sigma$ confidence level as inferred from the MCMC optimization. The parameters are provided for the cluster dark matter halo (dPIE DMH) and the BCG halo (dPIE BCG). $\Delta$R.A. and $\Delta$Decl. are defined in relation to the center of the BCG, located at R.A.=3:26:49.96 and Decl.=-0:43:51.47. Position angles are measured north of west, and the ellipticity $\epsilon$ is defined as $(a^2-b^2)/(a^2+b^2)$. $r_{cut}$ is fixed to 1000kpc for the cluster dark matter halo.}
	\label{tab.ms0326model}
	\begin{tabular}{lccccccc} 
		\hline
		Object & $\Delta$R.A. (") & $\Delta$Decl. (") & $\epsilon$ & $\theta$ ($^\circ$) & $r_{\mathrm{core}}$ (") & $r_{\mathrm{cut}}$  (") & $\sigma$ (km $s^{-1}$) \\
		\hline
dPIE DMH    & -0.92  & -0.75 & 0.23 & 134.57 & 42.17 & 1000. & 924.55 \\
dPIE BCG    & 17.56 & -20.75 & - & -  & -  & - & 189.44 \\
		\hline
	\end{tabular}
\end{table*}

\begin{table*}
	\centering
	\caption{Parameters for the best-fit lens model of \msthree. Error bars correspond to $1\sigma$ confidence level as inferred from the MCMC optimization. The parameters are provided for the cluster dark matter halo (dPIE DMH) and the BCG halo (dPIE BCG). $\Delta$R.A. and $\Delta$Decl. are defined in relation to the center of the BCG, located at R.A.=21:40:15.16 and Decl.=-23:39:40.09. Position angles are measured north of west, and the ellipticity $\epsilon$ is defined as $(a^2-b^2)/(a^2+b^2)$. $r_{cut}$ is fixed to 1000kpc for Halo 1.}
	\label{tab.ms1427model}
	\begin{tabular}{lccccccc} 
		\hline
		Object & $\Delta$R.A. (") & $\Delta$Decl. (") & $\epsilon$ & $\theta$ ($^\circ$) & $r_{\mathrm{core}}$ (") & $r_{\mathrm{cut}}$  (") & $\sigma$ (km $s^{-1}$) \\
		\hline
dPIE DMH    & 0.24  & 0.012 & 0.35 & 148.77 & 13.51 & 1000.& 784.81 \\
dPIE BCG    & 0  & -0.018 & - & - & - & - & 277.92 \\
		\hline
	\end{tabular}
\end{table*}

\begin{table*}
	\centering
	\caption{Statistical values derived from the MCMC fitting for the models of each cluster presented in Section~\ref{sec:massmodels}. The columns show the likelihood, log$~L$; the rms deviation from the predicted positions of the multiple images from their observed positions in the image plane, $rms$, and the reduced $\chi^2$.}
	\label{tab.modelstats}
	\begin{tabular}{lcccc} 
		\hline
		Cluster & log~$L$ &  $rms$ & $\chi^2$ \\
		\hline
A383    & 18.73  & 0.61 & 35.76 \\
MS2137    & 16.01  & 0.70 & 14.22 \\
\mstwo    & 11.05 & 0.85 & 23.08 \\
\msthree    & 7.41  & 0.61 & 0.83 \\
MACSJ0417    & 47.66  & 0.49 & 44.29 \\
MACSJ0949    & 80.17  & 0.41 & 17.58 \\
		\hline
	\end{tabular}
\end{table*}

\begin{table*}
	\centering
	\caption{Measured redshifts in \abell. Column 1 is the ID of the source. Columns 2 and 3 are the R.A. and Decl. in degrees (J2000). Column 4 is the redshift of the source. Column 5 is the quality flag (QF) assigned to the redshift. The QF scales in quality from largest to smallest; a flag value of 3 indicates that we have high confidence in the value for the redshift, whereas a flag value of 1 indicates that we have low confidence in the value for the redshift.}
	\label{tab.a383MUSE}
	\begin{tabular}[t]{lcccc} 
		\hline
		ID & R.A. & Decl. & $z$ & QF \\
		\hline
1    &	42.01870947    &	-3.53513970   &	0.196 & 3\\        
3    &	42.01413524    &	-3.53591378   &	0.195 & 3\\
4    &	42.00714291    &	-3.53595815   &	0.412 & 3\\
7    &	42.00745227    &	-3.53755278   &	0.960 & 3\\
11   &	42.01410348    &	-3.52926656   &	0.188 & 3\\        
18   &	42.15254730    &	-3.53287324   &	1.010 & 3\\        
50   &	42.02032160    &	-3.53682543   &	1.011 & 3\\
57   &	42.01432510    &	-3.52883110   &	1.011  & 3\\
61   &	42.00947290    &	-3.52844810   &	4.633 & 3\\
70   &	42.01179557    &	-3.53284732   &	1.009 & 3\\
79   &	42.01001298    &	-3.53386725   &	1.505 & 3\\
80   &	42.01003680    &	-3.53068700   &	4.634   & 3\\
83   &	42.01280980    &	-3.52573640   &	4.636 & 3\\
180  &	42.00583728    &	-3.53475354   &	0.928 & 3\\
208  &	42.01527605    &	-3.53287594   &	0.190 & 3\\
242  &	42.00980114    &	-3.53086269   &	1.010 & 3\\
313  &	42.02047584    &	-3.53382125   &	1.092 & 3\\
394  &	42.01915633    &	-3.53195987   &	0.373 & 3\\
410  &	42.01286378    &	-3.53353040   &	1.010 & 3\\
451  &	42.00810054    &	-3.53202013   &	0.186 & 3\\
493  &	42.01919295    &	-3.53294806   &	6.031 & 3\\
501  &	42.01157428    &	-3.52974105   &	0.186 & 3\\
737  &	42.00958438    &	-3.53049744   &	1.010 & 3\\
786  &	42.01923866    &	-3.52626206   &	0.194 & 3\\
1007 &	42.01156363    &	-3.52457326   &	0.656 & 3\\
1014 &	42.01560756    &	-3.52639592   &	0.190 & 3\\
1103 &	42.01513734    &	-3.52105068   &	0.195 & 3\\
1150 &	42.02093902    &	-3.52339690   &	0.890 & 3\\
1183 &	42.01893922    &	-3.52259989   &	0.094  & 3\\
1353 &	42.01953604    &	-3.52455550   &	0.937 & 3\\
1482 &	42.01180784    &	-3.52436324   &	0.656   & 3\\
1578 &	42.01363994    &	-3.52635524   &	6.032 & 3\\
\vdots & \vdots & \vdots  & \vdots  & \vdots   \\ 
\hline
\end{tabular}
\begin{tabular}[t]{lcccc}
		\hline
		ID & R.A. & Decl. & $z$ & QF \\
		\hline
\vdots & \vdots & \vdots  & \vdots  & \vdots   \\ 
1601 &	42.02098120    &	-3.52583488   &	0.824 & 3\\
1741 &	42.01678438    &	-3.52648827   &	0.373 & 3\\
17   &	42.01396670    &	-3.53270560   &	0.279 & 2\\        
135  &	42.02019771    &	-3.53514563   &	0.137 & 2\\
154  &	42.01317449    &	-3.53516389   &	0.191 & 2\\
329  &	42.00860414    &	-3.53278628   &	0.188 & 2\\
570  &	42.00709600    &	-3.53079003   &	0.764 & 2\\ 
584  &	42.02237334    &	-3.53141806   &	0.182 & 2 \\
1141 &	42.00768973    &	-3.52797491   &	1.276 & 2\\
1223 &	42.02037639    &	-3.52286060   & 0.186   & 2\\
1377 &	42.01522849    &	-3.52388490   &	0.000     & 2\\
1407 &	42.00603086    &	-3.52378361   &	4.943 & 2\\
1531 &	42.01179860    &	-3.52578326   &	0.188  & 2\\
1582 &	42.01767138    &	-3.52618242   &	0.191 & 2\\
1953 &	42.01700107    &	-3.52165509   &	0.593 & 2\\
5    &	42.01549764    &	-3.53712308   &	0.189  & 1\\
77   &	42.01967533    &	-3.53680155   &	0.187  & 1\\
82   &	42.01976480    &	-3.52540490	  & 6.259   & 1\\
110  &  42.01184380    &	-3.53511810   &	2.681   & 1\\
112  &	42.02161230    &	-3.52512230   &	5.033   & 1\\
262  &  42.01449268    &	-3.53497007   &	0.000     & 1\\
270  &	42.01340539    &	-3.53364156   &	0.159  & 1\\
316  &	42.00696144    &	-3.53265637   &	0.185  & 1\\
457  &	42.02281348    &	-3.53305797   &	0.000     & 1\\
746  &	42.00876902    &	-3.52981517   &	0.195   & 1\\
793  &	42.00807525    &	-3.52938858   &	0.183  & 1\\
859  &	42.01083523    &	-3.52946545   &	0.187  & 1\\
914  &	42.00876656    &	-3.52927118   &	1.559 & 1\\
988  &	42.01139524    &	-3.52879504   &	0.187   & 1\\
996  &	42.01737083    &	-3.52816428   &	1.580 & 1\\
1526 &	42.01010281    &	-3.52584537   &	0.192  & 1\\
1923 &	42.01348284    &	-3.52118876   &	0.184  & 1\\
\hline
	\end{tabular}
\end{table*}

\begin{table*}
	\centering
	\caption{Measured redshifts in \ms. Column 1 is the ID of the source. Columns 2 and 3 are the R.A. and Decl. in degrees (J2000). Column 4 is the redshift of the source. Column 5 is the quality flag (QF) assigned to the redshift. The QF scales in quality from largest to smallest; a flag value of 3 indicates that we have high confidence in the value for the redshift, whereas a flag value of 1 indicates that we have low confidence in the value for the redshift.}
	\label{tab.ms2137MUSE}
	\begin{tabular}[t]{lccccc} 
		\hline
		ID & R.A. & Decl. & $z$ & QF \\
		\hline
2     &	325.0612502    &    -23.66751981  & 0.163 & 3\\
8     &	325.0629407    &    -23.65686360  & 1.495 & 3\\
51    &	325.0591929    &    -23.66169910  & 1.494 & 3\\
90    &	325.0650339    &    -23.66763583  & 1.496  & 3\\
93    & 325.0587208    &    -23.66655827  &	-4.0E-5 & 3\\
143   &	325.0631642    &	-23.66114100  & 0.314 & 3\\
265   &	325.0658341    &	-23.66693218  & 1.191   & 3\\
305   &	325.0600151    &	-23.66525997  & 0.317 & 3\\
432   &	325.0624174    &	-23.65700482  & 1.495 & 3\\
554   &	325.0717926    &	-23.66225558  & 0.539   & 3\\
557   &	325.0554316    &	-23.65875178  & 0.974 & 3\\
564   &	325.0658002    &    -23.66232491  & 0.123   & 3 \\
575   &	325.0652998    &	-23.66272321  & 3.086 & 3\\
579   &	325.0659165    &	-23.65859792  & 0.313 & 3 \\
594   &	325.0660545    &	-23.65719416  & 0.323  & 3\\
717   &	325.0626655    &	-23.66044969  & 0.313 & 3 \\
796   &	325.0625407    &	-23.66026792  & 0.313 & 3 \\
804   &	325.0608755    &	-23.65919716  & 0.314 & 3 \\
899   &	325.0603961    &	-23.65473848  & 1.265 & 3 \\
978   &	325.0641418    &	-23.65707151  & 1.495 & 3\\
982   &	325.0655182    &	-23.65546970  & 0.311 & 3\\
984   &	325.0577362    &	-23.65544930  & 0.313 & 3\\
1022  &	325.0573735    &	-23.65524711  & 3.086 & 3\\
1047  &	325.0631649    &	-23.65676262  & 1.495 & 3\\
1085  &	325.0646960    &	-23.65727918  & 1.495 & 3\\
1098  &	325.0681213    &	-23.65712822  & 0.317 & 3\\
1213  &	325.0627779    &	-23.65956801  & 1.191 & 3\\
1437  &	325.0607713    &	-23.65322025  & 0.317 & 3 \\
1467  &	325.0555769    &	-23.65364726  & 0.281 & 3 \\
1477  &	325.0710864    &	-23.65315590  & 0.315 & 3\\
\vdots & \vdots & \vdots  & \vdots  & \vdots   \\ 
\hline
\end{tabular}
\begin{tabular}[t]{lcccc}
		\hline
		ID & R.A. & Decl. & $z$ & QF \\
		\hline
\vdots & \vdots & \vdots  & \vdots  & \vdots   \\ 
263   &	325.0614965    &    -23.66630868  & 0.000     & 2\\
290   &	325.0612886    &	-23.66452073  & 0.316 & 2\\
307   &	325.0596807    &	-23.66552547  & 0.314 & 2 \\
616   &	325.0577885    &	-23.66006395  & 0.315 & 2\\
689   &	325.0631461    &	-23.65977012  & 1.496   & 2\\
1129  &	325.0649292    &	-23.65706087  & 1.492 & 2\\
1     &	325.0557350    &    -23.66728671  & 0.999   & 1\\
4     &	325.0613087    &	-23.66945031  & 0.310    & 1\\
63    &	325.0581008    &    -23.65941150  & 5.509   & 1\\
69    & 325.0639614    &	-23.66911864  & 0.000     & 1\\
166   &	325.0574263    &	-23.66726300  & 0.281 & 1 \\
221   &	325.0605980    &	-23.66653599  & 0.328 & 1 \\
224   &	325.0592894    &	-23.66707065  & 1.543   & 1\\
242   &	325.0622637    &	-23.66447865  & 0.323   & 1\\
269   &	325.0616751    &	-23.66637744  & 2.459  & 1 \\
299   &	325.0544046    &	-23.66521798  & 0.222   & 1\\
318   &	325.0596921    &    -23.66584184  & 0.000     & 1\\
431   &	325.0621176    &	-23.66431571  & 0.639 & 1 \\
503   &	325.0536364    &	-23.66107318  & 0.316   & 1\\
580   &	325.0633467    &	-23.66185286  & 0.316   & 1\\
644   &	325.0648793    &	-23.66182589  & 0.937 & 1\\
654   &	325.0642144    &    -23.66216598  & 0.000     & 1 \\
657   &	325.0602975    &    -23.66170037  & 0.000     & 1\\
805   &	325.0543628    &	-23.66046275  & 0.311   & 1\\
814   &	325.0653077    &	-23.66082818  & 2.744 & 1\\
868   &	325.0544333    &    -23.66021964  & 0.000     & 1\\
869   &	325.0580939    &	-23.66019602  & 0.318   & 1\\
1128  &	325.0631522    &	-23.65935947  & 1.496   & 1\\
1261  &	325.0699090    &    -23.65869802  & 0.000     & 1\\
1389  &	325.0630390    &    -23.66044158  & 0.000     & 1\\

		\hline
	\end{tabular}
\end{table*}

\begin{table*}
	\centering
	\caption{Measured redshifts in \mstwo. Column 1 is the ID of the source. Columns 2 and 3 are the R.A. and Decl. in degrees (J2000). Column 4 is the redshift of the source. Column 5 is the quality flag (QF) assigned to the redshift. The QF scales in quality from largest to smallest; a flag value of 3 indicates that we have high confidence in the value for the redshift, whereas a flag value of 1 indicates that we have low confidence in the value for the redshift.}
	\label{tab.ms0326MUSE}
	\begin{tabular}[t]{lccccc} 
		\hline
		ID & R.A. & Decl. & $z$ & QF \\
		\hline
1 	   & 51.70806634   &	-0.73727226   &	0.440 & 3\\
13     & 51.70810324   &    -0.73102918   & 0.448 & 3\\
15	   & 51.71595908   &    -0.72429963   & 0.058 & 3\\
16	   & 51.71636815   &    -0.73054060	  & 1.431 & 3\\
18	   & 51.70415754   &	-0.72469514   &	0.000		& 3\\
20	   & 51.71355394   &    -0.72488020   & 1.247 & 3\\
21	   & 51.71485312   &	-0.72508425   &	0.449 & 3\\
26	   & 51.70869202   &	-0.72576593   &	0.447 & 3\\
29     & 51.70178080   &    -0.73721435   & 1.179   & 3\\
32	   & 51.70734694   &	-0.72652196   &	0.443 & 3\\
37	   & 51.70639179   &	-0.72749385   &	0.453 & 3\\
42	   & 51.71263975   &	-0.72735037   &	0.452 & 3\\
46	   & 51.70401058   &	-0.72803930   &	0.438	& 3\\
50	   & 51.71014411   &	-0.72838957   &	0.449 & 3\\
52	   & 51.70845413   &    -0.72881622   & 0.000	 	& 3\\
54	   & 51.70380841   &    -0.72941315   & 1.248 & 3\\
55     & 51.70662959   &    -0.72941321   & 0.232 & 3\\
57	   & 51.71376910   &    -0.72767990	  & 0.776 & 3\\
61	   & 51.70590437   &    -0.73047397   & 0.446 & 3\\
62	   & 51.70549223   &    -0.73047924   & 1.248 & 3\\
68	   & 51.71396070   &	-0.73078547   &	0.458 & 3\\
69	   & 51.71124680   &    -0.73372370	  & 0.494 & 3\\
74	   & 51.70481740   &    -0.73595180	  & 5.880 & 3\\
76	   & 51.71123314   &	-0.73155115   &	0.455 & 3\\
$78_M$   & 51.70655724   &	-0.73162210   &	0.000	    & 3\\
$78_P$   & 51.70691830   & 	-0.72437400	  &	4.980	& 3\\
81	   & 51.71233054   &    -0.73199644   & 1.098 & 3\\
84	   & 51.70674370   &    -0.73699870	  & 5.879	& 3\\
86     & 51.70180090   &    -0.73108310	  & 5.878 & 3\\
88	   & 51.71547523   &    -0.73306727   & 0.356 & 3\\
103	   & 51.71046650   &    -0.73443810   & 1.248 & 3\\
108	   & 51.70190169   &	-0.73491512   &	0.448 & 3\\
\vdots & \vdots & \vdots  & \vdots  & \vdots   \\ 
\hline
\end{tabular}
\begin{tabular}[t]{lcccc}
		\hline
		ID & R.A. & Decl. & $z$ & QF \\
		\hline
\vdots & \vdots & \vdots  & \vdots  & \vdots   \\ 
113	   & 51.71150751   &	-0.73442642   &	0.446 & 3\\
116	   & 51.71221832   &    -0.73567774   & 0.804 & 3\\
121	   & 51.7005703	   &    -0.73653835	  & 0.444	& 3 \\
122	   & 51.70323641   &    -0.73677771   & 0.325 & 3\\
138	   & 51.70211192   &	-0.73705436   &	1.272 & 3\\
139	   & 51.71652064   &    -0.73709426   & 0.804 & 3\\
152	   & 51.70378826   &    -0.73798640   & 1.181 & 3\\
160	   & 51.71135863   &	-0.72256970   &	0.453	& 3 \\
161	   & 51.71602364   &	-0.72303510   &	0.441 & 3\\
5	   & 51.71293750   &	-0.72474400   &	1.145	& 2\\
23	   & 51.70883990   &	-0.73340260   &	0.248 & 2\\
34	   & 51.70202020   &	-0.72244300   &	4.788 & 2\\
39	   & 51.70925786   &	-0.72748144   &	0.455 & 2\\
41	   & 51.71549023   &	-0.72772794   &	0.458 & 2\\
44	   & 51.71477132   &	-0.72779148   &	3.235	& 2 \\
60	   & 51.70565583   &	-0.73065953   &	0.453	& 2 \\
85	   & 51.70532607   &	-0.73234988   &	3.755	& 2\\
94	   & 51.70138800   &    -0.73333498	  & 0.414	& 2\\
155	   & 51.70649017   &	-0.73827485   &	0.450 & 2\\
17	   & 51.70943969   &	-0.72450109   &	0.447	& 1\\
35	   & 51.71171314   &	-0.72674022   &	1.248 & 1\\
47	   & 51.70971234   &	-0.72803052   &	0.452	& 1 \\
53	   & 51.70219623   &	-0.72919887   &	0.440	& 1\\
62	   & 51.70046960   &	-0.72423605   &	4.012 & 1\\
68	   & 51.70504950   &	-0.72801140   &	3.755	& 1 \\
77	   & 51.70667121   &	-0.73145698   &	0.000		& 1\\
86	   & 51.70681907   &	-0.73258985   &	0.443	& 1 \\
87	   & 51.71079159   &	-0.73274959   &	0.434	& 1 \\
90	   & 51.70686528   &	-0.73301888   &	0.446 & 1\\
105	   & 51.70739184   &	-0.73472983   &	0.448	& 1 \\
118	   & 51.70982389   &	-0.73588872   &	0.436	& 1 \\
145	   & 51.70424116   &	-0.73770541   &	0.441	& 1 \\
		\hline
	\end{tabular}
\end{table*}

\begin{table*}
	\centering
	\caption{Measured redshifts in \msthree. Column 1 is the ID of the source. Columns 2 and 3 are the R.A. and Decl. in degrees (J2000). Column 4 is the redshift of the source. Column 5 is the quality flag (QF) assigned to the redshift. The QF scales in quality from largest to smallest; a flag value of 3 indicates that we have high confidence in the value for the redshift, whereas a flag value of 1 indicates that we have low confidence in the value for the redshift.}
	\label{tab.ms1427MUSE}
	\begin{tabular}[t]{lccccc} 
		\hline
		ID & R.A. & Decl. & $z$ & QF \\
		\hline
25	  & 216.91619884	& -25.35766428	& 0.317 &	3 \\
26	  & 216.92217694	& -25.35378681	& 0.000 &	3 \\
31	  & 216.90642465	& -25.35723329	& 1.236   &	3 \\
32	  & 216.91803386	& -25.35737324	& 0.000    &	3 \\
36	  & 216.92494939	& -25.35576190	& 0.232   &	3 \\
40	  & 216.91399381	& -25.35600037	& 0.325   &	3 \\
41	  & 216.91008760	& -25.35561308	& 0.313  &	3 \\
42	  & 216.91088341	& -25.35678064	& 0.883  &	3 \\
50	  & 216.91485845	& -25.35028410	& 0.884  &	3 \\
54	  & 216.91193940	& -25.35514541	& 0.319  &	3 \\
64	  & 216.91063525	& -25.35363405	& 1.236  &	3 \\
75	  & 216.91544327	& -25.34620637	& 0.662  &	3 \\
84	  & 216.91035987	& -25.34988187	& 0.309  &	3 \\
88	  & 216.91447505	& -25.35061860	& 0.318  &	3 \\
93	  & 216.91584053	& -25.35056782	& 0.312  &	3 \\
94	  & 216.91691114	& -25.35060176	& 0.319  &	3 \\
99	  & 216.91571850	& -25.35135379	& 0.000    &	3 \\
101	  & 216.92028056	& -25.35105752	& 0.320  &	3 \\
108	  & 216.91500246	& -25.35168478	& 0.313   &	3 \\
114	  & 216.91434723	& -25.35196763	& 0.207   &	3 \\
117	  & 216.91959666	& -25.35234976	& 0.321  &	3 \\
118	  & 216.91229301	& -25.35263420	& 0.325  &	3 \\
122   & 216.91602417	& -25.35293113  & 0.322  &	3 \\
131	  & 216.91432235	& -25.35339907	& 0.313  &	3 \\
132	  & 216.91465475	& -25.35352760	& 0.001   &	3 \\
133	  & 216.91264386	& -25.35362011	& 0.883   &	3 \\
148	  & 216.91775549	& -25.35420614	& 0.312  &	3 \\
156	  & 216.90745832	& -25.35435398	& 4.323   &	3 \\
162	  & 216.90934813	& -25.35488005	& 0.324  &	3 \\
167	  & 216.90690018	& -25.35510802	& 0.313   &	3 \\
180	  & 216.90829896	& -25.35540450	& 0.914  &	3 \\
191	  & 216.91571984	& -25.34906701	& 0.317  &	3 \\
193	  & 216.90811293	& -25.34955274	& 0.328  &	3 \\
195	  & 216.91512646	& -25.34888806	& 0.316  &	3 \\
196	  & 216.91563524	& -25.34970724	& 0.316  &	3 \\
207	  & 216.91835520	& -25.34833384	& 0.325  &	3 \\
217	  & 216.92013294	& -25.34813976	& 0.320  &	3 \\
219	  & 216.90701865	& -25.34757320	& 0.695  &	3 \\
225	  & 216.90930004	& -25.34377121	& 0.322   &	3 \\
226	  & 216.91207270	& -25.34651632	& 0.663  &	3 \\
234	  & 216.90966217	& -25.34486568	& 0.856  &	3 \\
\vdots & \vdots & \vdots  & \vdots  & \vdots   \\ 
\hline
\end{tabular}
\begin{tabular}[t]{lcccc}
		\hline
		ID & R.A. & Decl. & $z$ & QF \\
		\hline
\vdots & \vdots & \vdots  & \vdots  & \vdots   \\ 
241	  & 216.92268855	& -25.34577380	& 0.322  &	3 \\
251	  & 216.91990633	& -25.34613558	& 1.239   &	3 \\
1	  & 216.91549900	& -25.35909370	& 0.232   &	2 \\
56	  & 216.91033590	& -25.35151940	& 0.813  &	2 \\
68	  & 216.91654650	& -25.35620440	& 6.043  &	2 \\
72	  & 216.91702200	& -25.35768770	& 3.961  &	2 \\
73	  & 216.91282910	& -25.34527350	& 6.042  &	2 \\
81	  & 216.92296584	& -25.34974507	& 0.000    &	2 \\
96	  & 216.91911505	& -25.35069157	& 0.780    &	2 \\
102	  & 216.92348540	& -25.35112651	& 0.437   &	2 \\
103	  & 216.90951026	& -25.35120070	& 0.315  &	2 \\
138	  & 216.92258186	& -25.34430751	& 1.119  &	2 \\
199	  & 216.91479272	& -25.34911416	& 0.000    &	2 \\
218	  & 216.90615033	& -25.34387282	& 0.915  &	2 \\
220	  & 216.91306431	& -25.34663774	& 0.315  &	2 \\
2	  & 216.90763630	& -25.35900540	& 0.326   &	1 \\	
22	  & 216.91948152	& -25.35840617	& 0.346   &	1 \\
34	  & 216.91844686	& -25.35742171	& 2.055   &	1 \\
35	  & 216.91288324	& -25.35744786	& 0.321   &	1 \\
53	  & 216.92419220	& -25.35758180	& 3.711   &	1 \\
57	  & 216.91773293	& -25.35578650	& 0.347   &	1 \\
58	  & 216.91795080	& -25.34248430	& 5.987   &	1 \\
74	  & 216.92431599	& -25.35543093	& 0.231  &	1 \\
77	  & 216.91119860	& -25.34467890	& 4.474   &	1 \\
121	  & 216.91576496	& -25.35285091	& 1.381   &	1 \\
140	  & 216.90804218	& -25.35371079	& 0.907  &	1 \\
141	  & 216.91940086	& -25.35381755	& 0.309   &	1 \\
144	  & 216.91105772	& -25.35395397	& 0.398   &	1 \\
164	  & 216.92323652	& -25.35479505	& 0.317   &	1 \\
169	  & 216.92067451	& -25.35487997	& 0.575   &	1 \\
170	  & 216.91419559	& -25.34846838	& 4.019   &	1 \\
172	  & 216.90966979	& -25.35508015	& 0.318   &	1 \\
173	  & 216.92047912	& -25.35515232	& 0.575   &	1 \\
176	  & 216.90934137	& -25.35527508	& 0.328   &	1 \\
185	  & 216.90949896	& -25.34656741	& 1.208  &	1 \\
201	  & 216.91901817	& -25.34852172	& 1.040    &	1 \\
208	  & 216.91808783	& -25.34834611	& 3.349   &	1 \\
213	  & 216.90558929	& -25.34828333	& 0.913  &	1 \\
232	  & 216.90673382	& -25.34649694	& 0.783  &	1 \\
248	  & 216.91130553	& -25.34609367	& 0.327  &	1 \\
254	  & 216.91156687	& -25.34619676	& 0.319   &	1 \\
		\hline
	\end{tabular}
\end{table*}


\bsp	
\label{lastpage}
\end{document}